\documentclass[reviewcopy]{elsarticle}

\usepackage[reviewcopy]{adndt}
\usepackage{longtable}
\usepackage{multirow}

\usepackage{amsmath}

\usepackage{amssymb}

\biboptions{square,sort&compress}
\bibpunct[]{[}{]}{,}{n}{}{;}
\begin{document}

\begin{frontmatter}

\journal{Atomic Data and Nuclear Data Tables}


\title{An Update of B(E2) Evaluation for $0_{1}^{+} \rightarrow 2_{1}^{+}$ Transitions in Even-Even Nuclei near N$\sim$Z$\sim$28}

  \author[One]{B. Pritychenko\corref{cor1}}
  \ead{E-mail: pritychenko@bnl.gov}

  \author[Two]{J. Choquette}
  \author[Three] {M. Horoi}
  \author[Two]{B. Karamy} 
  \author[Two]{B. Singh}

  \cortext[cor1]{Corresponding author.}

  \address[One]{National Nuclear Data Center, Brookhaven National Laboratory, Upton, NY 11973-5000, USA}

  \address[Two]{Department of Physics \& Astronomy, McMaster University, Hamilton, Ontario L8S 4M1, Canada}

  \address[Three]{Department of Physics, Central Michigan University,
Mount Pleasant, MI 48859, USA}

\date{16.12.2002} 

\begin{abstract}  
An update of the B(E2)$\uparrow$ evaluation for even-even Cr, Fe, Ni and Zn nuclei has been presented. 
The current update is a continuation of S. Raman {\it et al.}'s work on B(E2)$\uparrow$ values and was motivated 
by a large number of new measurements.  
It extends the previous evaluation from 20 to 38 nuclei and includes a comprehensive shell-model analysis. 
Evaluation policies for analysis of experimental data have been discussed. 
Future plans for a complete B(E2)$\uparrow$ evaluation of even-even nuclei are outlined.
\end{abstract}

\end{frontmatter}




\newpage

\tableofcontents
\listofDtables
\listofDfigures
\vskip5pc


\section{Introduction}
 
Quadrupole collectivities (reduced electric quadrupole transition rates), or B(E2)$\uparrow$ values, play an important role in nuclear physics and  are in 
high demand for nuclear model calculations. Originally, these values were compiled by S. Raman {\it et al.} at the Oak Ridge 
Nuclear Data Project \cite{87Ram,01Ram}. Presently, this work continues within the U.S. Nuclear Data Program (USNDP). In 2005, the Brookhaven B(E2)$\uparrow$ 
project website ({\it http://www.nndc.bnl.gov/be2}) was successfully launched  \cite{08Pri}; this website currently contains  up-to-date compilation of 
B(E2;$0_{1}^{+} \rightarrow 2_{1}^{+}$) experimental results and evaluated values from Raman {\it et al.} \cite{01Ram} that are widely used by scientists.

With an advent of rare isotope facilities \cite{06MSU} the whole nuclear landscape has been changing dramatically. These  facilities have been producing rare 
nuclei far from the valley of stability at an increasing rate and providing researchers with unprecedented opportunities to study their properties. 
In many cases, B(E2)$\uparrow$ values and energies of the low-lying states have been studied for the first time. Large amounts of new data, especially for the 
A$\leq$100 region, require a new evaluation of quadrupole collectivities for proper interpretation and analysis of the newly-obtained values. 
A renewed interest in N$\sim$Z$\sim$28 B(E2) values was expressed by participants of the International Nuclear Physics Conference (INPC 2010) in 
Vancouver ({\it http://inpc2010.triumf.ca}) \cite{10Van}.

To answer the need for a new B(E2)$\uparrow$ evaluation and seek comments from the research community, new evaluations of Cr, Fe, Ni and Zn isotopes 
have been completed. The complete evaluation of B(E2)$\uparrow$ for 
all even-even nuclei would follow thereafter based on our experience and the feedback from the users.

\section{B(E2)$\uparrow$ Evaluation Policies}

The current evaluation represents an update of B(E2)$\uparrow$ in $e^{2}b^{2}$, lifetimes ($\tau$) in ps 
and deformation parameter ($\beta_{2}$) values for Cr, Fe, Ni and Zn nuclei. These values are mutually related:
\begin{equation}
\label{eq0}
\tau =  40.81 \times 10^{13} E^{-5}_{\gamma} [B(E2)\uparrow / e^{2}b^{2}]^{-1} (1+\alpha_{T})^{-1} 
\end{equation}
\begin{equation}
\label{eq1}
\beta_{2} = (4\pi / 3 ZR_0^2)[B(E2)\uparrow/ e^{2}]^{1/2}
\end{equation}
where E$_{\gamma}$ and $\alpha_{T}$ are the $\gamma$-ray energy in $keV$ and the total conversion coefficient, respectively, and R$_{0}^{2}$ = (1.2$\times$10$^{-13}$A$^{1/3}$cm)$^2$. To introduce an additional measure of collectivity for nuclear excitations, Weisskopf units (W.u.) are added. 
Transition quadrupole moment values $Q_0$ in $b$ were not included in the current evaluation 
\begin{equation}
\label{eq2}
Q_{0} = [16 \pi B(E2)\uparrow / 5 e^2]^{1/2}
\end{equation}
All measured values can be grouped using three classes of experimental techniques:
\begin{itemize}
\item Model-independent or traditional types of measurements \cite{01Ram}: transmission Doppler-shift attenuation lifetime (TDSA), recoil distance (RDM), delayed coincidences (TCS), low-energy and intermediate-energy Coulomb  excitation (CE) and nuclear resonance fluorescence ($\gamma$,$\gamma^{\prime}$).
\item Low model-dependent: electron scattering (E,E$^{\prime}$), hyperfine splitting.
\item Model-dependent: inelastic scattering of light and heavy ions (IN-EL).
\end{itemize}

\subsection{Nuclear Databases}
Nuclear Science References (NSR) \cite{NSR}, Evaluated Nuclear Structure Data File (ENSDF) \cite{90ENSDF,96ENSDF} and 
Experimental Unevaluated Nuclear Data List (XUNDL) \cite{XUNDL} databases played a crucial role in this project. 
 A short description of the databases is presented below.

The NSR database \cite{NSR} is the most comprehensive source of low- and intermediate-energy nuclear physics bibliographical information,  
containing more than 200,000 articles since the beginning of nuclear science. It consists of primary (journals) and secondary (proceedings, lab reports, theses, private communications) references. 
The main goal of the NSR is to provide bookmarks for experimental and 
theoretical articles in nuclear science using keywords. NSR keywords are assigned to articles that contain results on atomic nuclei and masses, nuclear decays, 
nuclear reactions and other properties. Keywords are also used to build author and subject indexes, which allow users to search for articles by subject (Coulomb excitation, $\sigma$, B(E2), T$_{1/2}$, ...) or author. 
This database is updated on a weekly basis and serves as a source of bibliographical information for the ENSDF database.

The ENSDF database \cite{90ENSDF,96ENSDF} contains evaluated nuclear structure and decay data in a standard format. An international network of evaluators \cite{nsdd} 
contributes to the database. For each nuclide, all known experimental data used to deduce nuclear structure information are included. 
Each type of experiment is presented as a separate dataset. In addition, there is a dataset of ``adopted" level 
and $\gamma$-ray transition properties, which represent the evaluator's determination of the best values for these properties, 
based on all available experimental data. Information in the database is regularly 
updated and published in Elsevier Nuclear Data Sheets journal. Due to the large scope of the database, evaluation updates 
are often conducted on a 6-12 year basis.

The XUNDL database \cite{XUNDL} contains compiled experimental nuclear structure data in the ``ENSDF" format. In general, the information 
in a given XUNDL dataset comes from a single journal article, or from a set of closely-related articles by one group of authors and later used in the ENSDF evaluations.

We primarily used NSR and XUNDL databases for the experimental data search. These searches were verified using the ENSDF database, 
previous evaluation of S. Raman {\it et al.} \cite{01Ram} and references from the original experimental papers.

\subsection{B(E2)$\uparrow$ Evaluation Procedure}
This evaluation is based on the analysis of results from 114 primary and 13 secondary references published prior to April 2011. 
The evaluation procedure for the derivation of adopted (recommended) B(E2)$\uparrow$ values is presented below:
\begin{itemize}
\item Compile a list of experimental B(E2)$\uparrow$,$\downarrow$ or W.u., $\tau$ and $\beta_2$ values as reported in the original papers without any changes or modifications. Reported values depend on the measured quantities and are deduced from experimental data in the offline analysis.
\item Convert experimental  values into B(E2)$\uparrow$ in  $e^{2}b^{2}$. 
\item Analyze B(E2)$\uparrow$ values. In a few of the older results, where uncertainties were not quoted by the authors, we have taken
the values as adopted by Raman {\it et al.} \cite{01Ram}. The minimum uncertainty assigned to a datum in the averaging procedure was 5\%. 
The experimental values listed in Table 3, however, show the uncertainties as quoted by the authors.
\item Round uncertainties to two significant digits.
\item Follow the procedure by Raman {\it et al.} \cite{01Ram} for asymmetric uncertainties: consider the upper and lower bounds, 
extract the mean of the two values and assign an uncertainty so that the value overlaps the two bounds.
\item Deduce B(E2)$\uparrow$ recommended values using model-independent or traditional, combined (model-independent and low model-dependent) and model-dependent data sets with AveTools software package \cite{10Ave} using the selected data sets.
\end{itemize}

\section{Adopted B(E2)$\uparrow$ values}

The recommended values from the current project for Cr, Fe and Ni isotopes are shown in Table \ref{table1}. Compared  
to the previous evaluation of S. Raman {\it et al.} \cite{01Ram}, it includes 18 new recommended values for $^{46,56,58,60,62}$Cr, $^{50,52,62,64,66}$Fe, $^{54,70,74}$Ni and $^{72,74,76,78,80}$Zn. 
A complementary analysis of the two evaluations is presented below.

In the current evaluation, we  used the latest AveTools averaging procedures \cite{10Ave}, Band-Raman  calculation of Internal conversion coefficients ($\alpha_{T}$) \cite{08Bri} and presently available data.
The program AveTools \cite{10Ave} combines limitation of relative statistical weight (LWM), normalized residual (NRM) and Rajeval technique (RT) statistical 
methods  \cite{92Ave,Lwei} to calculate averages of experimental data with uncertainties. In the present work, we start with the weighted mean values followed by LWM, NRM and RT by accepting reduced $\chi^{2}$$<$2 as a reasonable fit for available data sets.
Previously, S. Raman {\it et al.} \cite{01Ram} used an averaging procedure based on the inverse of the 
quoted uncertainties, while current evaluation uses statistical methods  based on the inverse squared value of the quoted uncertainties.

The Band-Raman method \cite{08Bri} was used in this work, while the previous evaluation \cite{01Ram} employed the internal conversion coefficients code (ICCDF)  \cite{93Icc}. 
This code incorporates the Dirac-Fock atomic model with the exchange interaction between atomic
electrons  and the free electron receding to infinity during the conversion process. In the Cr, Fe, Ni and Zn-region with low Z-values and relatively high 2$^{+}_{1}\rightarrow$ 0$^{+}_{1}$ transition energies, the
total E2 conversion coefficients are relatively small ($\alpha_{T} <$ 0.002) to substantially affect the adopted values. A complementary comparison between the present model-independent and the previous evaluation 
adopted values for $^{54}$Cr and $^{54}$Fe, where no new data were added, 
shows good agreement. Consequently, the differences between the current work and  S. Raman {\it et al.} \cite{01Ram} evaluation are mainly due to  the addition of new  
experimental results. 

We recommend using model-independent or traditional B(E2)$\uparrow$ adopted values as the most reliable. If a model-independent value is not available,  
a combined value should be used. Finally, a model-dependent value can be used if no other values are available. 
This is consistent with the previous evaluation of Raman {\it et al.} \cite{01Ram}, 
who treated data as follows: ``However, our adopted B(E2)$\uparrow$ values are based only on the traditional types of
measurements because these are more direct and involve essentially model-independent analyses." 
The new recommended values are interpreted within the scope of large-scale shell-model calculations which are presented in the following sections.

\section{Shell Model Calculations}\label{chapter4}
The $2^+$ excitation energies and B(E2) for  $0_{1}^{+} \rightarrow 2_{1}^{+}$ transitions have been calculated 
in the $pf$-shell valence space using the GXPF1A effective
interaction \cite{gxpf1a}. 
GXPF1A is a refinement of the original GXPF1A Hamiltonian \cite{gxpf1}, which was obtained starting with the G-matrix for the Bonn-C two-body potential
and by further fine-tuning its matrix elements to describe the energies of about 700 selected states of $pf$-shell nuclei. 
The GXPF1 Hamiltonian does not describe very well the 2$^+$ state in $^{54}$Ti (N=34); as a result, five of its matrix elements were changed to fix this discrepancy, leading to the
GXPF1A Hamiltonian \cite{gxpf1a}. 
GXPF1A predicts the 2$^+$ state in $^{58}$Cr (N=36) at higher energy than that seen in the experimental data, but
one would not expect to get reliable energies when the number of neutrons is close to the limits of the $pf$-shell (N=40). 
Results for Cr, Fe, Ni and Zn nuclei using the ``canonical" effective charges, 0.5e for neutrons and 1.5e for protons, 
are shown in Table \ref{table2} and Fig. 1-4. 

The missing values for $^{66}$Fe, $^{68,70,74}$Ni, $^{72,74,76,78,80}$Zn are due to the limitations of the 
valence space and of the GXPF1A effective interaction.  
In the $pf$-shell one cannot have more than 20 valence protons and 20 valence neutrons on top of the $^{40}$Ca core. 
Even when N is too close to the limit (N=40) the results are not reliable, due to the increasing importance of 
the intruder states ($g_{9/2}$), which create an ``island of inversion" \cite{90Wa}. 
Therefore, for the heavy isotopes of Ni and Zn we performed shell-model calculations in the 
$f_{5/2}, p_{3/2}, p_{1/2}, g_{9/2}$ valence space using the JUN45 effective interaction \cite{09Ho}.
Results obtained using the effective charges recommended in Ref. \cite{09Ho} for this model space,
$e_p$=1.5e and $e_n$=1.1e,  are shown in Table \ref{table2}.

\begin{figure}[!htb]
\includegraphics[height=10cm, angle=0]{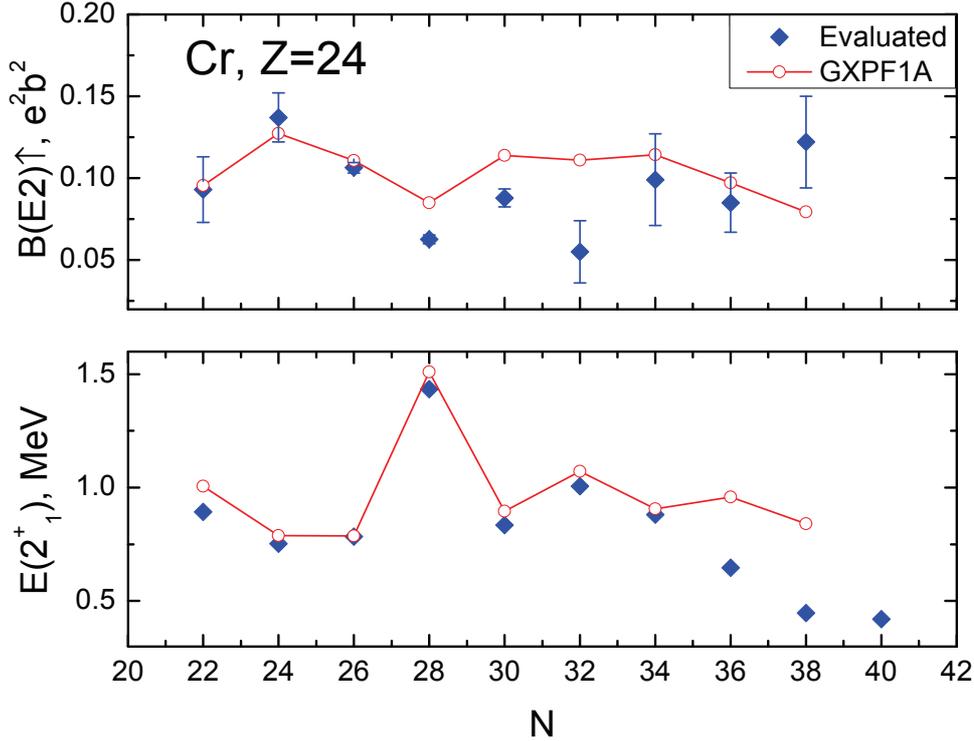}
\caption{Shell model calculated and evaluated E($2^{+}_{1}$) and B(E2)$\uparrow$ values for Cr nuclei.}
\label{fig1}
\end{figure}
\begin{figure}[!htb]
\includegraphics[height=10cm, angle=0]{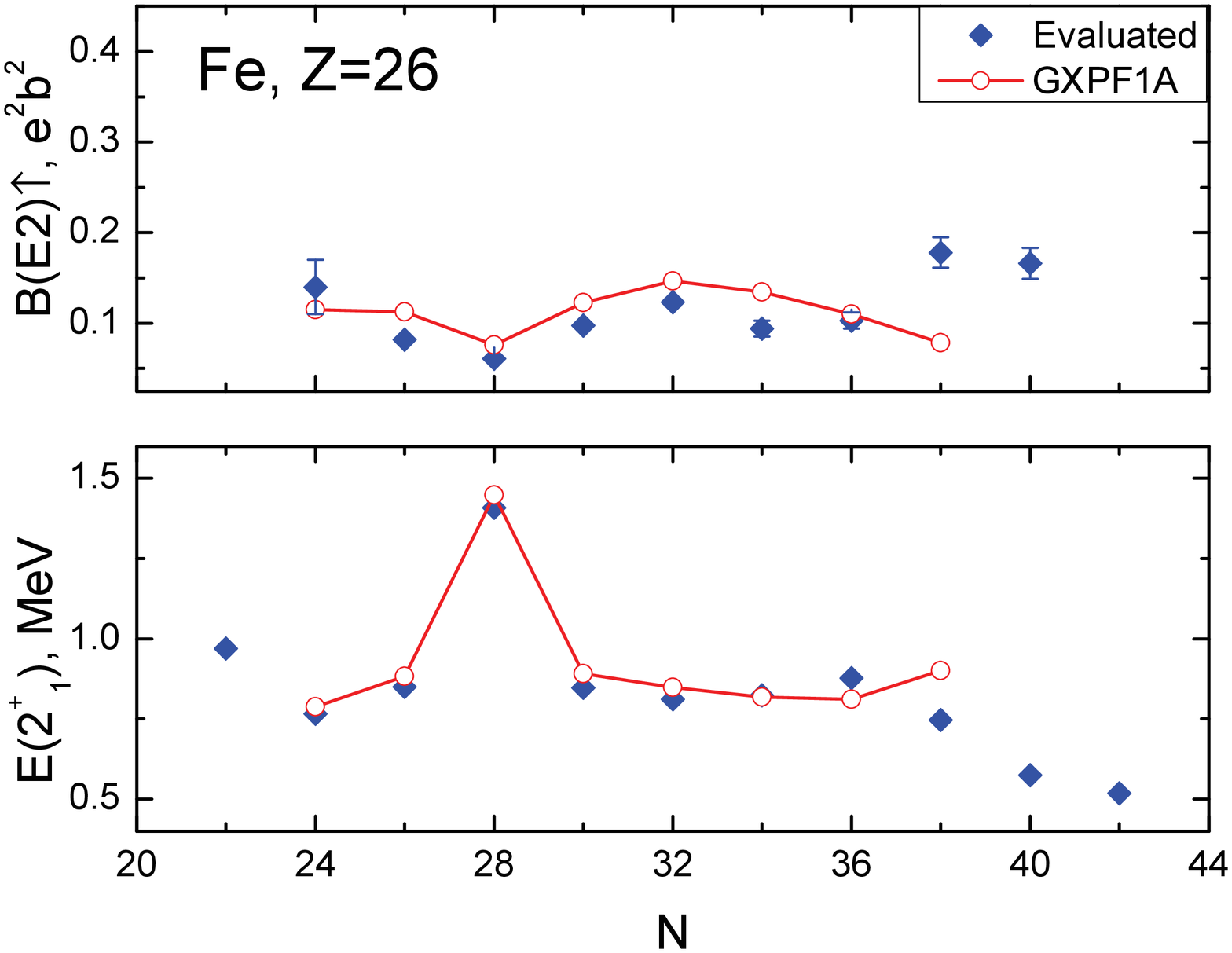}
\caption{Shell model calculated and evaluated  E($2^{+}_{1}$) and B(E2)$\uparrow$ values for Fe nuclei.}
\label{fig2}
\end{figure} 
\begin{figure}[!htb]
\includegraphics[height=10cm, angle=0]{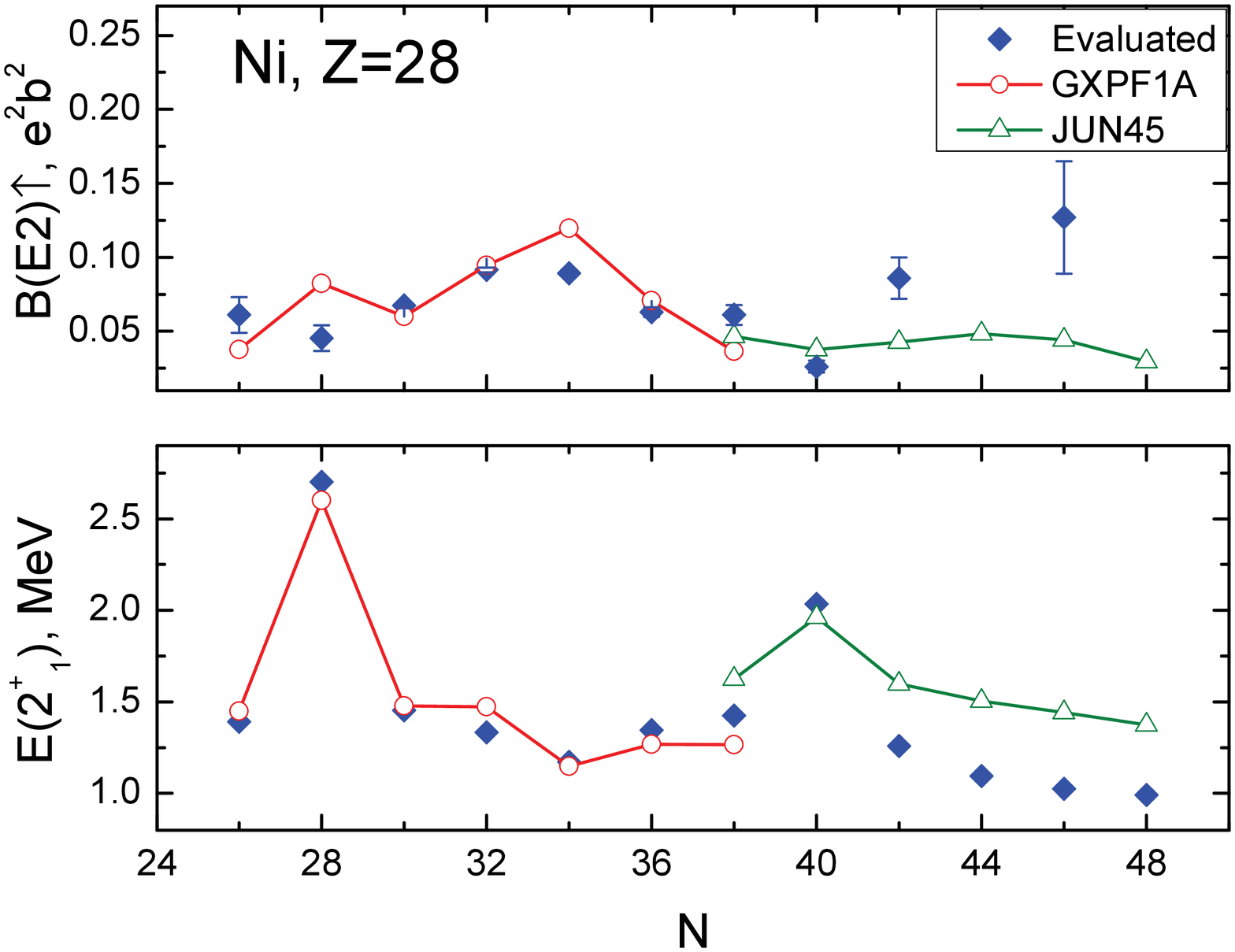}
\caption{Shell model calculated and evaluated  E($2^{+}_{1}$) and B(E2)$\uparrow$ values for Ni nuclei.}
\label{fig3}
\end{figure}
\begin{figure}[!htb]
\includegraphics[height=10cm, angle=0]{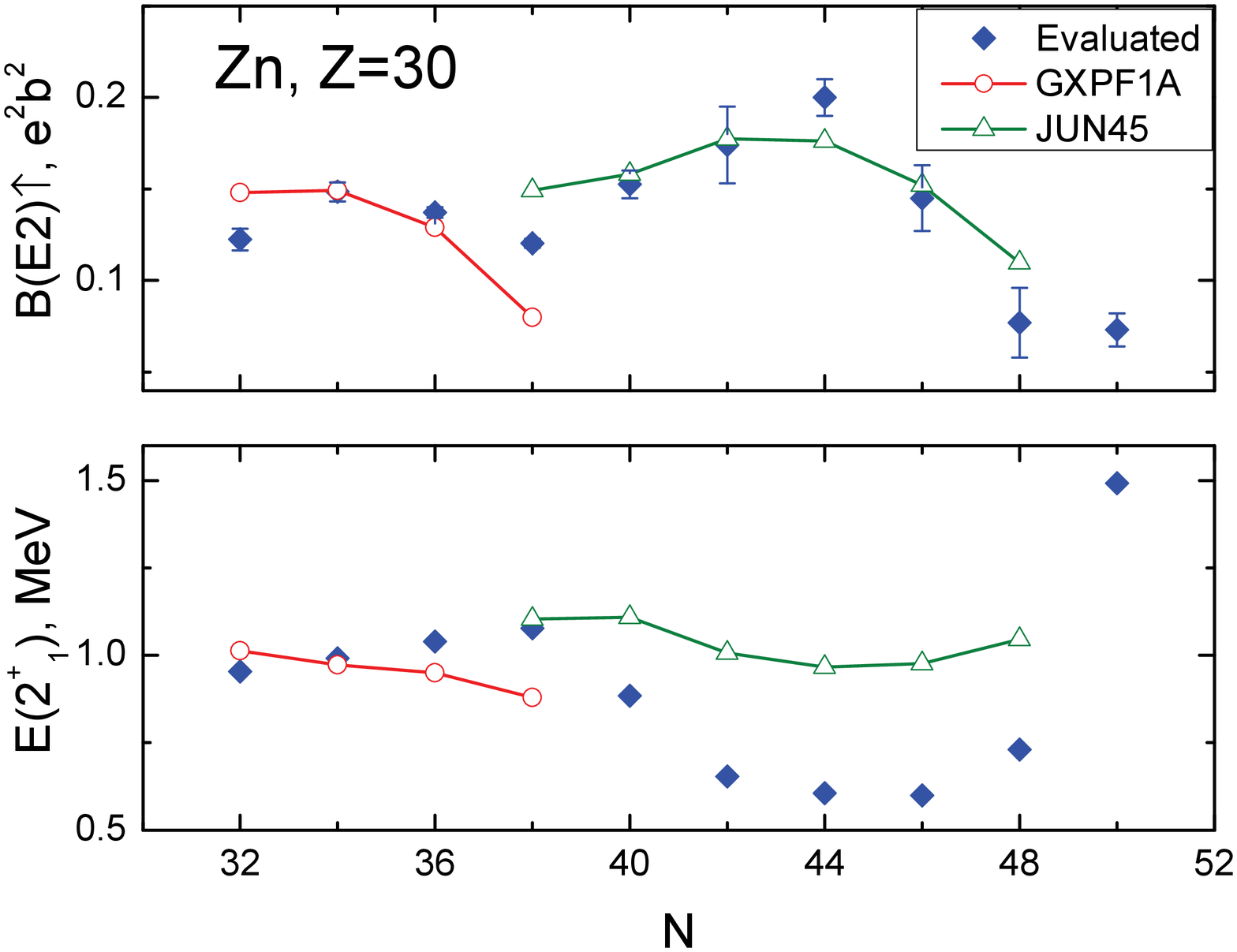}
\caption{Shell model calculated and evaluated  E($2^{+}_{1}$) and B(E2)$\uparrow$ values for Zn nuclei.}
\label{fig4}
\end{figure} 

\section{Experimental B(E2)$\uparrow$ values}
 Experimental values of B(E2)$\uparrow$, $\tau$ and $\beta_{2}$ are shown in Table \ref{table3}. 
 To create a more comprehensive picture for each experiment we extended the scope of the previous work of S. Raman {\it et al.} \cite{01Ram} 
 and  included target, beam, beam energy and a flag for the Coulomb barrier height into compilation.   
 A short review of the most recent experimental results used for the new evaluation is presented below.

\subsection{$^{46}$Cr, $^{50}$Fe, $^{54}$Ni:} 
To complete systematics in the $N = Z = 28$ region, B(E2)$\uparrow$ values of 0.093(20), 0.140(30) and 0.059(17) e$^2$b$^2$  have been 
reported in intermediate-energy Coulomb excitation of $^{46}$Cr, $^{50}$Fe, $^{54}$Ni \cite{2005Ya26}, respectively.

\subsection{$^{56,58}$Cr:}
Relativistic Coulomb excitation B(E2)$\downarrow$ values of $^{56,58}$Cr are  8.7(3.0) and 14.8(4.2) W.u., respectively, 
have been measured by the RISING collaboration \cite{2005Bu29}. These results agree with the shell-model calculation based 
on GXPF1A and GXPF1 effective interactions \cite{gxpf1a,gxpf1}.

\subsection{$^{60,62}$Cr:}
Deformation length and quadrupole deformation parameter have been measured in inelastic scattering of Chromium on Hydrogen \cite{2009Ao01} and provide evidence for enhanced collectivity in chromium nuclei.

\subsection{$^{52}$Fe:}
Intermediate-energy Coulomb excitation measurements at Michigan State University (MSU) \cite{2004Yu07}  have produced a B(E2)$\uparrow$ 
value of 0.082(10) e$^2$b$^2$. The increase in  E2 strength with respect to the even-mass neighbor $^{54}$Fe agrees 
with shell-model calculations as the magic number N=28 is approached. 

\subsection{$^{62,64,66}$Fe:}
The $^{62,64}$Fe lifetimes of 7.4(9) and 7.4(26) ps \cite{2010Lj01} were reported  by the GANIL group using 
the recoil-distance Doppler shift method after multinucleon transfer reactions in inverse kinematics.
These results corroborate recent MSU measurements of 8.0(10), 10.3(10), 39.0(40) ps for $^{62,64,66}$Fe \cite{2011Ro02}, respectively. 
The deduced E2 strengths demonstrate the enhanced collectivity of the neutron-rich Fe isotopes up to $N=40$. Note that both use a plunger method.

\subsection{$^{70}$Ni:}
The reduced transition probability B(E2)$\uparrow$ of  0.086(14) e$^2$b$^2$ \cite{2006Pe13} 
for the neutron-rich  $^{70}$Ni nucleus has been measured by Coulomb excitation in a $^{208}$Pb target at intermediate energy. 
The current B(E2)$\uparrow$ value for $^{70}$Ni is unexpectedly large, which may indicate that neutrons added above $N=40$ strongly polarize the $Z=28$ proton core.

\subsection{$^{74}$Ni:}
The deformation length and quadrupole deformation parameter have been measured in inelastic scattering of $^{74}$Ni on Hydrogen \cite{2010Ao01}. 
Results of this experiment indicate that the magic character of $Z=28$ or $N=50$ is weakened in $^{74}$Ni.

\subsection{$^{72}$Zn:}
The reduced transition probabilities B(E2)$\uparrow$ of  0.174(21) e$^2$b$^2$ \cite{2002Le17} 
for  $^{72}$Zn nucleus has been measured by Coulomb excitation at intermediate energy.  
This result is consistent with the expectations derived from the neighboring nucleus $^{73}$Zn and 
indicates that the behavior of E2 strengths around the $N = 40$ sub-shell closure in Zn is very different from the Ni isotopic chains.

\subsection{$^{74}$Zn:}
A lifetime of 27.6(43) ps was recently reported in the recoil distance Doppler-shift measurement at GANIL \cite{2011Ni03}.  
This result agrees well with the previous B(E2)$\uparrow$ values of  0.201(16) e$^2$b$^2$ and 0.204(15) e$^2$b$^2$ measured at  REX-ISOLDE and GANIL \cite{2007Va20,2006Pe13}, respectively.

\subsection{$^{76,78,80}$Zn:}
The reduced transition probabilities B(E2)$\uparrow$ of  0.145(18), 0.077(19) and 0.073(9) e$^2$b$^2$  
for $^{76,78,80}$Zn have been reported by the REX-ISOLDE group \cite{2007Va20,2009Va01}  using a low-energy Coulomb excitation. 
The present data indicate a need for large-scale shell-model calculations.

\section{Conclusion \& Outlook}
An updated B(E2)$\uparrow$ evaluation of even-even Cr, Fe, Ni and Zn isotopes has been performed under the auspices of the USNDP with an intention to 
update B(E2)$\uparrow$ values and collect nuclear data user feedback. 
It is a continuation of S. Raman work on transition 
probabilities from the ground to the first-excited 2$^+$ state of even-even nuclides \cite{87Ram,01Ram}. 
The update is based on all published data prior to April 2011  and includes new experimental B(E2) values for 33 out of 38 nuclei. 
It extends evaluated data in the N$\sim$Z$\sim$28 region 
from 20 to 38 nuclei. These results are compared with large-scale shell-model calculations. 
Evaluations of quadrupole collectivities for all nuclides, grouped by Z region, will follow, accommodating user feedback based on this paper.

\section{Acknowledgments}
The authors are grateful to Prof. J. Cameron (McMaster University) and V. Unferth (Viterbo University) for  productive discussions and  careful reading of the manuscript 
and useful suggestions, respectively. This work was funded by the Office of Nuclear Physics, Office of Science of the U.S. Department of Energy, 
under Contract No. DE-AC02-98CH10886 with Brookhaven Science Associates, LLC. 
Work at McMaster University was also supported by DOE and NSERC of Canada. MH acknowledges support from DOE grant DE-FC02-09ER41584 (UNEDF SciDAC Collaboration).

\newpage

\TableExplanation



\bigskip
\renewcommand{\arraystretch}{1.0}

\section*{Table 1.\label{tbl1te} Adopted (recommended) B(E2)$\uparrow$-, $\tau$- and $\beta_2$-values for Cr, Fe, Ni and Zn isotopes.}
\begin{tabular*}{0.95\textwidth}{@{}@{\extracolsep{\fill}}lp{5.5in}@{}}
\multicolumn{2}{p{0.95\textwidth}}{Throughout this table,
	bracketed  numbers refer to the uncertainties in the last
	digits of the quoted values; no star character, $^{*}$ and $^{**}$ correspond to model-independent, combined and model-dependent values, respectively.}\\

Nuclide	& The even $Z$, even $N$ nuclide studied\\

$E$(level)
	& Energy of the first excited 2$^+$ state in keV either from an
	ENSDF evaluation or from current literature\\

$B(E2)\!\!\uparrow$
	& Reduced electric quadrupole transition rate for the ground
	state to $2^+$ state transition in units of ${\textrm{e}^2}$b$^2$\\

$\tau$	& Mean lifetime of the state in ps. The relation between $\tau$ and B(E2)$\uparrow$ is given as \\
	& \quad$\tau =
  40.81\times10^{13}E^{-5}_{\gamma}[B(E2)\!\!\uparrow/{\textrm{e}^2}\textrm{b}^2]^{-1} 
		(1+\alpha_{T})^{-1}$, where E$_{\gamma}$ is the $\gamma$-ray energy and $\alpha_{T}$ is the
total conversion coefficient \\

$\beta_{2}$
	& Quadrupole deformation parameter deduced from B(E2)$\uparrow$\\
	& \quad\begin{array}[t]{l@{=}l}
	  \beta_{2} & (4\pi/3 ZR_0^2)[B(E2)\!\!\uparrow/{\textrm{e}^2}]^{1/2},\
    \textrm{where}\\ 
		R_{0}^{2} & (1.2\times10^{-13}A^{1/3}\textrm{cm})^2\\
		          &0.0144 A^{2/3}\textrm{b}
	  \end{array}\\
$B(E2)\!\!\uparrow$ \cite{01Ram}
	& Previous value of reduced electric quadrupole transition rate for the ground
	state to $2^+$ state transition in units of ${\textrm{e}^2}$b$^2$\\
\end{tabular*}
\label{tableI}

\section*{Table 2.\label{tbl2te} Calculated E(2$_1^+$)-, B(E2$\uparrow$)-values for Cr, Fe, Ni and Zn isotopes.}
\begin{tabular*}{0.95\textwidth}{@{}@{\extracolsep{\fill}}lp{5.5in}@{}}
Nuclide	& The even $Z$, even $N$ nuclide studied\\

GXPF1A  effective interaction \cite{gxpf1a} 
	& From chapter \ref{chapter4} shell-model calculations \\
$E$(level)
	& GXPF1A energy value of the first excited 2$^+$ state in MeV \\

$B(E2)\!\!\uparrow$
	& GXPF1A reduced electric quadrupole transition rate value for the ground
	state to $2^+$ state transition in units of ${\textrm{e}^2}$b$^2$\\
	
JUN45 effective interaction \cite{09Ho} 
	& From chapter \ref{chapter4} shell-model calculations \\
$E$(level)
	& JUN45 energy value  of the first excited 2$^+$ state in MeV \\

$B(E2)\!\!\uparrow$
	& JUN45 reduced electric quadrupole transition rate value for the ground
	state to $2^+$ state transition in units of ${\textrm{e}^2}$b$^2$\\	
\end{tabular*}
\label{tableII}

\section*{Table 3.\label{tbl3te} Experimental B(E2$\uparrow$)-, $\tau$- and $\beta_{2}$-values in Cr, Fe, Ni and Zn isotopes.}
\begin{tabular*}{0.95\textwidth}{@{}@{\extracolsep{\fill}}lp{5.5in}@{}}
\multicolumn{2}{p{0.95\textwidth}}{Throughout this table,
	bracketed numbers refer to the uncertainties in the last
	digits of the quoted values. $^{s}$,$^{d}$ or * - Superseded, duplicate  or above the Coulomb barrier \cite{99Zag} experiments. 
	Beam energy units are in MeV or (A)-MeV/nucleon. NSR keynumbers \cite{NSR} are shown in the reference column.}\\

Nuclide	& The even $Z$, even $N$ nuclide studied\\

$B(E2)\!\!\uparrow$
	& Reduced electric quadrupole transition rate for the ground
	state to $2^+$ state transition in units of ${\textrm{e}^2}$b$^2$\\

$\tau$	& Mean lifetime of the state in ps\\

$\beta_{2}$
	& Quadrupole deformation parameter\\

Target & Target nuclide\\
Beam & Incident beam \\
Energy & Incident beam energy\\
Method & CE: Coulomb excitation\\
       & CE$^{*}$: Coulomb excitation with beam energy above the Coulomb barrier\\
       & CE?: Coulomb excitation, incomplete information \\
       & TDSA: Transmission Doppler shift attenuation lifetime \\
       & TDSA$^{r}$: Rejected as an outlier \\
       & RDM: Measurement as a function of distance of the relative fraction of recoil nuclei which decay in a movable plunger \\
       & TCS: Observation, with fast electronics, of the delay between transitions in a cascade \\
       & $\gamma$,$\gamma^{\prime}$: Measurement of the nuclear resonance fluorescence cross section \\
       & E,E$^{\prime}$: Inelastic electron scattering\\
       & IN-EL: Inelastic scattering of light and heavy ions\\       
Reference & NSR database \cite{NSR} keynumber\\

\end{tabular*}
\label{tableIII}

\newpage
\datatables 

\setlength{\LTleft}{0pt}
\setlength{\LTright}{0pt} 


\setlength{\tabcolsep}{0.5\tabcolsep}

\renewcommand{\arraystretch}{1.0}

\footnotesize 

\begin{longtable}{@{\extracolsep\fill}cllllll@{}}
\caption{Adopted (recommended) B(E2)$\uparrow$-, $\tau$- and $\beta_2$-values for Cr, Fe, Ni and Zn isotopes. } \label{table1}
\textbf{Nuclide} & E$_{2^+_1}$  & \multicolumn{2}{c}{\textbf{B(E2)$\uparrow$} }  & \textbf{$\tau$ }  & \textbf{$\beta_{2}$} & \textbf{B(E2)$\uparrow$} \cite{01Ram}  \\
	& (keV)   & ($e^{2}b^{2}$) & (W.u.) & (ps) & & ($e^{2}b^{2}$) 	\\
\hline \hline
\endfirsthead
\caption[]{Adopted (recommended) B(E2)$\uparrow$-, $\tau$- and $\beta_2$-values for Cr, Fe, Ni and Zn isotopes (continued).}
\textbf{Nuclide} & E$_{2^+_1}$  & \multicolumn{2}{c}{\textbf{B(E2)$\uparrow$} }  & \textbf{$\tau$ }  & \textbf{$\beta_{2}$} & \textbf{B(E2)$\uparrow$} \cite{01Ram}  \\
	&  (keV)  & ($e^{2}b^{2}$) & (W.u.) & (ps) & & ($e^{2}b^{2}$) 	\\
\hline \hline
\endhead
\multirow{2}{*}{$^{46}$Cr}  &  \multirow{2}{*}{892.16(10)}  & 0.093(20) & 19.0(41) & 16.7(36) & 0.288(31) &  \\  
 &  &  &  &  &  &   \\\\ \hline\\
\multirow{2}{*}{$^{48}$Cr}  &  \multirow{2}{*}{752.19(11)}  & 0.137(15) & 26.4(29) & 12.4(14) & 0.340(19) & 0.136(21)   \\   
 &  &  &  &  &  &   \\\\ \hline\\
\multirow{2}{*}{$^{50}$Cr}  &  \multirow{2}{*}{783.30(9)}  & 0.1063(32) & 19.43(58) & 13.02(39) & 0.2912(32) & 0.108(6)  \\   
 &  & 0.1046(27)$^{*}$ & 19.12(49)$^{*}$ & 13.23(34)$^{*}$ &  &   \\\\ \hline\\
\multirow{2}{*}{$^{52}$Cr}  &  \multirow{2}{*}{1434.094(14)}  & 0.0627(27) & 10.88(47) & 1.076(46) & 0.2179(47) & 0.0660(30)  \\ 
  &  & 0.0650(13)$^{*}$ & 11.28(23)$^{*}$ & 1.035(21)$^{*}$ &  &   \\\\ \hline\\
\multirow{2}{*}{$^{54}$Cr}  &  \multirow{2}{*}{834.855(3)}  & 0.0879(55) & 14.50(91) & 11.45(72) & 0.2509(75) & 0.0870(40)  \\ 
  &  & 0.0910(45)$^{*}$ & 15.01(74)$^{*}$ & 11.06(55)$^{*}$ &  &   \\\\ \hline\\
\multirow{2}{*}{$^{56}$Cr}  &  \multirow{2}{*}{1006.61(20)}  & 0.055(19) & 8.7(30) & 7.1(25) & 0.195(34) &  \\  
 &  &  &  &  &  &   \\\\ \hline\\
\multirow{2}{*}{$^{58}$Cr}  &  \multirow{2}{*}{880.7(2)}  & 0.099(28) & 14.8(42) & 7.8(22) & 0.254(37) &  \\  
 &  &  &  &  &  &   \\\\ \hline\\
\multirow{2}{*}{$^{60}$Cr}  &  \multirow{2}{*}{646(1)}  &  &  &  & 0.23(3)$^{**}$ &  \\
   &  & 0.085(18)$^{**}$ & 12.3(27)$^{**}$ & 43(11)$^{**}$ &  &   \\\\ \hline\\
\multirow{2}{*}{$^{62}$Cr}  &  \multirow{2}{*}{447(4)}  &  &  &  & 0.27(3)$^{**}$ &  \\ 
  &  & 0.122(28)$^{**}$ & 16.7(38)$^{**}$ & 187(45)$^{**}$ &  &   \\\\ \hline\\
\multirow{2}{*}{$^{64}$Cr}  &  \multirow{2}{*}{420(7)}  &  &  &  &  &  \\ 
  &  &  &  &  &  &   \\\\ \hline\\
\multirow{2}{*}{$^{48}$Fe}  &  \multirow{2}{*}{969.5(5)}  &  &  &  &  &  \\ 
  &  &  &  &  &  &   \\\\ \hline\\
\multirow{2}{*}{$^{50}$Fe}  &  \multirow{2}{*}{765.0(10)}  &  &  &  & 0.308(33)$^{*}$ &  \\ 
  &  & 0.140(30)$^{*}$ & 25.6(55)$^{*}$ & 11.1(24)$^{*}$  &  &   \\\\ \hline\\
\multirow{2}{*}{$^{52}$Fe}  &  \multirow{2}{*}{849.45(10)}  &  &  &  & 0.230(14)$^{*}$ &  \\ 
  &  & 0.082(10)$^{*}$ & 14.2(18)$^{*}$ & 11.3(14)$^{*}$ &  &   \\\\ \hline\\
\multirow{2}{*}{$^{54}$Fe}  &  \multirow{2}{*}{1408.19(19)}  & 0.0608 (31) & 10.0(5) & 1.21(6) & 0.193(5) & 0.062(5)   \\ 
  &  & 0.0542(18)$^{*}$ & 8.94(30)$^{*}$ & 1.36(5)$^{*}$ &  &   \\\\ \hline\\
\multirow{2}{*}{$^{56}$Fe}  &  \multirow{2}{*}{846.776(5)}  & 0.0975(27) & 15.32(42) & 9.61(27) & 0.239(3) & 0.0980(40)  \\ 
  &  & 0.0970(22)$^{*}$ & 15.24(35)$^{*}$ & 9.66(22)$^{*}$ &  &   \\\\ \hline\\
\multirow{2}{*}{$^{58}$Fe}  &  \multirow{2}{*}{810.7662(20)}  & 0.123(4) & 18.4(6) & 9.55(31) & 0.261(5) & 0.1200(40)  \\ 
  &  & 0.0920(48)$^{*}$ & 13.8(7)$^{*}$ & 12.3(6)$^{*}$ &  &   \\\\ \hline\\
\multirow{2}{*}{$^{60}$Fe}  &  \multirow{2}{*}{823.63(15)}  & 0.0938(88) & 13.4(13) & 11.5(11) & 0.224(10) & 0.096(18)  \\ 
  &  &  &  &  &  &   \\\\ \hline\\
\multirow{2}{*}{$^{62}$Fe}  &  \multirow{2}{*}{876.8(3)}  & 0.1028(90) & 14.1(12) & 7.67(67) & 0.229(10) &  \\ 
  &  &  &  &  &  &   \\\\ \hline\\
\multirow{2}{*}{$^{64}$Fe}  &  \multirow{2}{*}{746.40(10)}  & 0.178(17) & 23.4(22) & 9.93(97) & 0.295(15) &  \\ 
  &  &  &  &  &  &   \\\\ \hline\\
\multirow{2}{*}{$^{66}$Fe}  &  \multirow{2}{*}{574.4(10)}  & 0.166(17) & 21.0(21) & 39.4(40) & 0.280(15) &  \\
 &  &  &  &  &  &   \\\\ \hline\\
\multirow{2}{*}{$^{68}$Fe}  &  \multirow{2}{*}{517(6)?}  &  &  &  &  &  \\ 
  &  &  &  &  &  &   \\\\ \hline\\
\multirow{2}{*}{$^{54}$Ni}  &  \multirow{2}{*}{1392.3(4)}  & 0.061(12) & 10.0(20) & 1.28(25) & 0.179(18) &  \\ 
  &  &  &  &  &  &   \\\\ \hline\\
\multirow{2}{*}{$^{56}$Ni}  &  \multirow{2}{*}{2700.6(7)}  & 0.0453(86) & 7.1(13) & 0.062(13) & 0.151(14) & 0.060(12)   \\ 
  &  & 0.0502(70)$^{*}$ & 7.9(11)$^{*}$ & 0.057(8)$^{*}$ &  &   \\\\ \hline\\
\multirow{2}{*}{$^{58}$Ni}  &  \multirow{2}{*}{1454.21(9)}  & 0.0673(17) & 10.09(25) & 0.933(24) & 0.1799(23) & 0.0695(20)  \\
   &  & 0.0646(17)$^{*}$ & 9.67(25)$^{*}$ & 0.971(25)$^{*}$ &  &   \\\\ \hline\\
\multirow{2}{*}{$^{60}$Ni}  &  \multirow{2}{*}{1332.518(5)}  & 0.0914(17) & 13.17(24) & 1.057(20) & 0.2055(19) & 0.0933(15)  \\ 
  &  & 0.0899(16)$^{*}$ & 12.89(23)$^{*}$ & 1.081(20)$^{*}$ &  &   \\\\ \hline\\
\multirow{2}{*}{$^{62}$Ni}  &  \multirow{2}{*}{1172.91(9)}  & 0.0893(21)   & 12.25(29) & 2.094(43) & 0.1982(23) & 0.0890(25)  \\ 
  &  & 0.0878(18)$^{*}$ & 12.05(25)$^{*}$ & 2.059(48)$^{*}$ &  &   \\\\ \hline\\
\multirow{2}{*}{$^{64}$Ni}  &  \multirow{2}{*}{1345.75(5)}  & 0.0629(32) & 8.27(42) & 1.47(7) & 0.1628(41) & 0.076(8)  \\ 
  &  & 0.0663(27)$^{*}$ & 8.72(35)$^{*}$ & 1.39(6)$^{*}$ &  &   \\\\ \hline\\
\multirow{2}{*}{$^{66}$Ni}  &  \multirow{2}{*}{1424.8(10)}  & 0.0611(67)  & 7.71(85) & 1.14(12) & 0.157(9) & 0.062(9)  \\
   &  &  &  &  &  &   \\\\ \hline\\
\multirow{2}{*}{$^{68}$Ni}  &  \multirow{2}{*}{2034.07(17)}  & 0.0260 (40) & 3.15(49) & 0.451(69) & 0.101(8) & 0.026(6)  \\ 
  &  &  &  &  &  &   \\\\ \hline\\
\multirow{2}{*}{$^{70}$Ni}  &  \multirow{2}{*}{1259.6(2)}  & 0.086(14) & 10.0(16) & 1.50(24) & 0.179(15) &  \\
   &  &  &  &  &  &   \\\\ \hline\\
\multirow{2}{*}{$^{72}$Ni}  &  \multirow{2}{*}{1096.0(20)}  &  &  &  &  &  \\ 
  &  &  &  &  &  &   \\\\ \hline\\
\multirow{2}{*}{$^{74}$Ni}  &  \multirow{2}{*}{1024(1)}  &  &  &  & 0.21(3)$^{**}$ &  \\ 
  &  & 0.127(38)$^{**}$ & 13.8(41)$^{**}$ & 2.86(85)$^{**}$ &  &   \\\\ \hline\\
\multirow{2}{*}{$^{76}$Ni}  &  \multirow{2}{*}{992(2)}  &  &  &  &  &  \\ 
  &  &  &  &  &  &   \\\\ \hline\\
\multirow{2}{*}{$^{60}$Zn}  &  \multirow{2}{*}{1003.9(2)}  &  &  &  &  &  \\ 
  &  &  &  &  &  &   \\\\ \hline\\
\multirow{2}{*}{$^{62}$Zn}  &  \multirow{2}{*}{954.0(4)}  & 0.1224(59) & 16.79(81) & 4.22(20) & 0.2166(52) & 0.124(9)   \\ 
  &  & 0.1224(59)$^{*}$ & 16.79(81)$^{*}$ & 4.22(20)$^{*}$ &  &   \\\\ \hline\\
\multirow{2}{*}{$^{64}$Zn}  &  \multirow{2}{*}{991.56(5)}  & 0.1484(52) & 19.52(68) & 2.87(10) & 0.2335(41) & 0.160(15)  \\
   &  & 0.1519(43)$^{*}$ & 19.98(57)$^{*}$ & 2.803(79)$^{*}$  &  &   \\\\ \hline\\
\multirow{2}{*}{$^{66}$Zn}  &  \multirow{2}{*}{1039.2279(21)}  & 0.1371(29) & 17.31(37) & 2.456(52) & 0.2198(24) & 0.135(10)  \\ 
  &  & 0.1389(31)$^{*}$ & 17.53(39)$^{*}$ & 2.424(54)$^{*}$ &  &   \\\\ \hline\\
\multirow{2}{*}{$^{68}$Zn}  &  \multirow{2}{*}{1077.37(4)}  & 0.1203(25) & 14.59(30) & 2.337(49) & 0.2019(21) & 0.124(15)  \\
   &  & 0.1198(28)$^{*}$ & 14.53(34)) & 2.347(55)$^{*}$ &  &   \\\\ \hline\\
\multirow{2}{*}{$^{70}$Zn}  &  \multirow{2}{*}{884.46(8)}  & 0.1525(75) & 17.80(88) & 4.93(24) & 0.2229(55) & 0.160(14)  \\
   &  & 0.169(14)$^{*}$ & 19.7(16)$^{*}$ & 4.45(37)$^{*}$ &  &   \\\\ \hline\\
\multirow{2}{*}{$^{72}$Zn}  &  \multirow{2}{*}{652.70(5)}  & 0.174(21) & 19.6(24) & 19.8(24) & 0.234(14) &  \\ 
  &  &  &  &  &  &   \\\\ \hline\\
\multirow{2}{*}{$^{74}$Zn}  &  \multirow{2}{*}{605.9(8)}  & 0.200(10) & 21.7(11) & 25.0(12) & 0.2460(62) &  \\ 
  &  &  &  &  &  &   \\\\ \hline\\
\multirow{2}{*}{$^{76}$Zn}  &  \multirow{2}{*}{598.68(10)}  & 0.145(18) & 15.2(19) & 36.6(45) & 0.206(12) &  \\ 
  &  &  &  &  &  &   \\\\ \hline\\
\multirow{2}{*}{$^{78}$Zn}  &  \multirow{2}{*}{730.2(4)}  & 0.077(19) & 7.8(19) & 25.5(63) & 0.147(18) &  \\ 
  &  &  &  &  &  &   \\\\ \hline\\
\multirow{2}{*}{$^{80}$Zn}  &  \multirow{2}{*}{1492(1)}  & 0.073(9) & 7.1(9) & 0.76(9) & 0.141(9) &  \\ 
  &  &  &  &  &  &   \\\\ \hline\\
\end{longtable}

\newpage
\begin{longtable}{@{\extracolsep\fill}cllll@{}}
\caption{Calculated E(2$^{+}_{1}$)-, B(E2)$\uparrow$-values for Cr, Fe, Ni and Zn isotopes.} \label{table2}
\textbf{Nuclide} & \multicolumn{2}{c}{GXPF1A effective interaction \cite{gxpf1a}} & \multicolumn{2}{c}{JUN45 effective interaction \cite{09Ho}} \\
 & \textbf{E(2$^{+}_{1}$) (MeV)} & \textbf{B(E2)$\uparrow$ ($e^{2}b^{2}$)} & \textbf{E(2$^{+}_{1}$) (MeV)} & \textbf{B(E2)$\uparrow$ ($e^{2}b^{2}$)} \\
\hline \hline
\endfirsthead
 \textbf{Nuclide} & \multicolumn{2}{c}{GXPF1A effective interaction \cite{gxpf1a}} & \multicolumn{2}{c}{JUN45 effective interaction \cite{09Ho}} \\
         & \textbf{E(2$^{+}_{1}$) (MeV)} & \textbf{B(E2)$\uparrow$ ($e^{2}b^{2}$)} & \textbf{E(2$^{+}_{1}$) (MeV)} & \textbf{B(E2)$\uparrow$ ($e^{2}b^{2}$)} \\
\hline \hline
\endhead
 $^{46}$Cr & 1.0054  &  0.0955 & & \\
 $^{48}$Cr & 0.7887  &  0.1273 & & \\
 $^{50}$Cr & 0.7872  &  0.1107 & & \\
 $^{52}$Cr & 1.5101  &  0.0849 & & \\
 $^{54}$Cr & 0.8949  &  0.1138 & & \\
 $^{56}$Cr & 1.0715  &  0.1109 & & \\
 $^{58}$Cr & 0.9062  &  0.1143 & & \\
 $^{60}$Cr & 0.958   &  0.0972 & & \\
 $^{62}$Cr & 0.840   &  0.0793 & & \\
 \hline \\
 $^{50}$Fe & 0.787   &  0.1151 & & \\
 $^{52}$Fe & 0.883   &  0.1124 & & \\
 $^{54}$Fe & 1.4483  &  0.0761 & & \\
 $^{56}$Fe & 0.8903  &  0.1228 & & \\
 $^{58}$Fe & 0.8478  &  0.1468 & & \\
 $^{60}$Fe & 0.8173  &  0.1345 & & \\
 $^{62}$Fe & 0.8114  &  0.1101 & & \\
 $^{64}$Fe & 0.9008  &  0.0784 & & \\
 $^{66}$Fe &         &         & & \\
\hline \\
 $^{54}$Ni & 1.448   &  0.0375 & & \\
 $^{56}$Ni & 2.599   &  0.0823 & & \\
 $^{58}$Ni & 1.478   &  0.0599 & & \\
 $^{60}$Ni & 1.474   &  0.0946 & & \\
 $^{62}$Ni & 1.149   &  0.1195 & & \\
 $^{64}$Ni & 1.268   &  0.0706 & & \\
 $^{66}$Ni & 1.265   &  0.0365 & 1.624  & 0.0464 \\
 $^{68}$Ni &         &         & 1.963  & 0.0376 \\
 $^{70}$Ni &         &         & 1.599  & 0.0427 \\
 $^{72}$Ni &         &         & 1.505  & 0.0483 \\
 $^{74}$Ni &         &         & 1.442  & 0.0440 \\
 $^{76}$Ni &         &         & 1.374  & 0.0296 \\
\hline \\
 $^{62}$Zn & 1.013   & 0.1479  & & \\
 $^{64}$Zn & 0.973   & 0.1492  & & \\
 $^{66}$Zn & 0.950   & 0.1290  & & \\
 $^{68}$Zn & 0.879   & 0.0799  & 1.104  & 0.1493 \\
 $^{70}$Zn &         &         & 1.109  & 0.1581 \\
 $^{72}$Zn &         &         & 1.007  & 0.1773 \\
 $^{74}$Zn &         &         & 0.966  & 0.1763 \\
 $^{76}$Zn &         &         & 0.976  & 0.1521 \\
 $^{78}$Zn &         &         & 1.045  & 0.1097 \\
 $^{80}$Zn &         &         & & \\ 
\hline \\
\end{longtable}

\newpage

\renewcommand{\arraystretch}{1.0}

\footnotesize
\begin{longtable}{@{\extracolsep\fill}cllllcccl@{}}
\caption{Experimental  B(E2$\uparrow$)-, $\tau$- and $\beta_{2}$-values in Cr, Fe, Ni and Zn isotopes.} \label{table3}
\textbf{Nuclide} & \textbf{B(E2)} ($e^{2}b^{2}$)  & \textbf{$\tau$ } (ps) &  \textbf{$\beta_{2}$} & \textbf{Target} & \textbf{Beam} & \textbf{Energy} (MeV) & \textbf{Method} & \textbf{Reference}\\
\hline \hline\\
\endfirsthead
\caption[]{Experimental  B(E2$\uparrow$)-, $\tau$- and $\beta_{2}$-values in Cr, Fe, Ni and Zn isotopes (continued).}
\textbf{Nuclide} & \textbf{B(E2)$\uparrow$} ($e^{2}b^{2}$)  & \textbf{$\tau$ } (ps) &  \textbf{$\beta_{2}$} & \textbf{Target} & \textbf{Beam} & \textbf{Energy} (MeV) & \textbf{Method} & \textbf{Reference}\\
\hline \hline\\
\endhead 

$^{46}$Cr & 0.093(20) &  &  & $^{208}$Pb & $^{46}$Cr & 44 A & CE* & \cite{2005Ya26} \\
$^{48}$Cr &  & 10.6(11) &  & $^{36}$Ar & $^{14}$N & 28-35  & RDM & \cite{1979Ek03} \\
$^{48}$Cr &  & 16.7(22) &  & $^{34}$S & $^{16}$O & 30-36  & RDM & \cite{1975Ha04} \\
$^{48}$Cr &  & 9.7(26) &  & $^{40}$Ca & $^{10}$B & 19-25  & RDM & \cite{1973Ku10} \\
$^{50}$Cr &  & 13.2(4) &  & $^{12}$C & $^{50}$Cr & 110-120  & CE* & \cite{2000Er01} \\
$^{50}$Cr & 0.093(5) &  &  & $^{50}$Cr & e- & 30-400  & EE$^{\prime}$ & \cite{1983Li02} \\
$^{50}$Cr & 0.102(5) &  &  & $^{50}$Cr & $^{32}$S & 62 & CE* & \cite{1975To06} \\
$^{50}$Cr &  & 12.6(21) &  & $^{40}$Ca & $^{16}$O & 47 & TDSA & \cite{1974Br04} \\
$^{50}$Cr &  & 12.1(12) &  & $^{40}$Ca & $^{12}$C & 28 & RDM & \cite{1973De09} \\
$^{50}$Cr &  & 10(2) &  & $^{52}$Cr & p & 31.4 & TDSA & \cite{1972Ra14} \\
$^{50}$Cr & 0.115(10) &  &  & $^{50}$Cr & $^{35}$Cl & 54 & CE & \cite{1972Ra14} \\
$^{50}$Cr & 0.092(10) &  &  & $^{50}$Cr & $^{16}$O/$^{35}$Cl & 21-79  & CE* & \cite{1971DaZM} \\
$^{50}$Cr & 0.115(8) &  &  & $^{50}$Cr & $^{4}$He &  & CE? & \cite{1961Mc18} \\
$^{50}$Cr & 0.15(3) &  &  & $^{50}$Cr & Ne & 23.2 & CE? & \cite{1960An09} \\
$^{52}$Cr &  & 1.13(3) &  & C & $^{52}$Cr & 110-120  & CE* & \cite{2000Er01} \\
$^{52}$Cr & 0.0632(40) &  &  & $^{52}$Cr & e- & 30-400  & EE$^{\prime}$ & \cite{1983Li02} \\
$^{52}$Cr & 0.0687(13) &  &  & $^{52}$Cr & $\gamma$ &  & GG & \cite{1981Ah02} \\
$^{52}$Cr & 0.080(8) &  &  & $^{52}$Cr & e- & 90, 120, 226  & EE$^{\prime}$ & \cite{1978Po04} \\
$^{52}$Cr & 0.0634(39) &  &  & $^{52}$Cr & e- & 40-110  & EE$^{\prime}$ & \cite{1976Li19} \\
$^{52}$Cr & 0.0660(30) &  &  & $^{52}$Cr & $^{32}$S & 60 & CE* & \cite{1975To06} \\
$^{52}$Cr & 0.076(8) &  &  & $^{52}$Cr & e- & 50,60,80,90  & EE$^{\prime}$ & \cite{1975DeXW} \\
$^{52}$Cr &  & 0.86(13) &  & $^{52}$Cr & $^{16}$O/$^{35}$Cl & 21-79  & TDSA & \cite{1972WaYZ} \\
$^{52}$Cr & 0.071(9) &  &  & $^{52}$Cr & e- & 150 & EE$^{\prime}$ & \cite{1971Pe11} \\
$^{52}$Cr &  & 0.99$^{+45}_{-25}$ &  & $^{51}$V & $^{3}$He & 11 & TDSA & \cite{1971Sp12} \\
$^{52}$Cr & 0.072(8) &  &  & $^{52}$Cr & $^{16}$O/$^{35}$Cl & 21-79  & CE* & \cite{1971DaZM} \\
$^{52}$Cr & 0.043(9) &  &  & $^{52}$Cr & $^{12}$C & 36.8 & CE* & \cite{1967Af03} \\
$^{52}$Cr & 0.048(2) &  &  & $^{52}$Cr & $^{16}$O  & 31-41  & CE* & \cite{1965Si02} \\
$^{52}$Cr & 0.0520(40) &  &  & $^{52}$Cr & e- & 150-180  & EE$^{\prime}$ & \cite{1964Be32} \\
$^{52}$Cr &  & 1.02(13) &  & $^{52}$Cr & $\gamma$ & 0.5-3  & GG & \cite{1964Bo22} \\
$^{52}$Cr & 0.073(7) &  &  & $^{52}$Cr & $^{4}$He &  & CE? & \cite{1961Mc18} \\
$^{52}$Cr & 0.060(15) &  &  & $^{52}$Cr & $^{16}$O & 39 & CE* & \cite{1960Ad01} \\
$^{52}$Cr & 0.062(12) &  &  & $^{52}$Cr & Ne & 23.2 & CE? & \cite{1960An09} \\
$^{52}$Cr &  & 0.8(2) &  & $^{52}$Cr & $\gamma$ & $<$2  & GG & \cite{1959Of14} \\
$^{54}$Cr & 0.095(5) &  &  & $^{54}$Cr & e- & 30-400  & EE$^{\prime}$ & \cite{1983Li02} \\
$^{54}$Cr & 0.0850(30) &  &  & $^{54}$Cr & $^{32}$S & 62 & CE* & \cite{1975To06} \\
$^{54}$Cr & 0.096(9) &  &  & $^{54}$Cr & $^{35}$Cl & 54 & CE & \cite{1970MiZQ} \\
$^{54}$Cr & 0.106(7) &  &  & $^{54}$Cr & $^{4}$He &  & CE? & \cite{1961Mc18} \\
$^{54}$Cr & 0.057(11) &  &  & $^{54}$Cr & $^{14}$N & 16.3, 26  & CE & \cite{1960An07} \\
$^{54}$Cr & 0.079(20) &  &  & $^{54}$Cr & $^{14}$N & 15.9-35  & CE & \cite{1959Al95} \\
$^{56}$Cr & 0.055(19) &  &  & $^{197}$Au & $^{56}$Cr & 100 A & CE* & \cite{2005Bu29} \\
$^{58}$Cr & 0.099(28) &  &  & $^{197}$Au & $^{58}$Cr & 100 A & CE* & \cite{2005Bu29} \\
$^{60}$Cr &  &  & 0.23(3) & p & $^{60}$Cr & 63 A & IN-EL & \cite{2009Ao01} \\
$^{62}$Cr &  &  & 0.27(3) & p & $^{62}$Cr & 63 A & IN-EL & \cite{2009Ao01} \\
$^{50}$Fe & 0.140(30) &  &  & Pb & $^{50}$Fe & 41 A & CE* & \cite{2005Ya26} \\
$^{52}$Fe & 0.082(10) &  &  & $^{197}$Au & $^{52}$Fe & 56.9 A & CE* & \cite{2004Yu07} \\
$^{54}$Fe & 0.0676(38) &  &  & $^{54}$Fe & $^{40}$Ca & 86* & CE & \cite{1981Le02} \\
$^{54}$Fe & 0.060(6) &  &  & $^{54}$Fe & e- & 50,60,80,90  & EE$^{\prime}$ & \cite{1975DeXW} \\
$^{54}$Fe &  & 1.10$^{+50}_{-32}$ &  & $^{54}$Fe & p & 10 & TDSA & \cite{1972Mo31} \\
$^{54}$Fe &  & 0.95(14) &  & $^{54}$Fe & $^{16}$O/$^{35}$Cl & 21-30; 60-79  & TDSA & \cite{1972WaYZ} \\
$^{54}$Fe & 0.0532(33) &  &  & $^{54}$Fe & e- & 150, 225  & EE$^{\prime}$ & \cite{1972Li28} \\
$^{54}$Fe & 0.0595(60) &  &  & $^{54}$Fe & $^{16}$O/$^{35}$Cl & 21-30; 60-79  & CE* & \cite{1971DaZM} \\
$^{54}$Fe & 0.061(14) &  &  & $^{54}$Fe & $^{12}$C & 36.8 & CE* & \cite{1967Af03} \\
$^{54}$Fe & 0.051(2) &  &  & $^{54}$Fe & $^{16}$O & 38.1 & CE* & \cite{1965Si02} \\
$^{54}$Fe & 0.0533(24) &  &  & $^{54}$Fe & e- & 150 & EE$^{\prime}$ & \cite{1962Be18} \\
$^{56}$Fe & 0.1022(55) &  &  & $^{12}$C/$^{56}$Fe & $^{52}$Cr & 22, 110-120  & CE* & \cite{1981Le02} \\
$^{56}$Fe &  & 7.9(12) &  & $^{51}$V & $^{7}$Li & 25 & RDM & \cite{1974Po15} \\
$^{56}$Fe & 0.111(6) &  &  & $^{56}$Fe & $^{4}$He/$^{16}$O & 7.5-30  & CE & \cite{1972Ca05} \\
$^{56}$Fe & 0.0970(20) &  &  & $^{56}$Fe & $^{32}$S & 65 & CE* & \cite{1972Le19} \\
$^{56}$Fe & 0.0678(48) &  &  & $^{56}$Fe & e- & 150, 225  & EE$^{\prime}$ & \cite{1972Li28} \\
$^{56}$Fe & 0.0945(45) &  &  & $^{56}$Fe & e- & 299.5 & EE$^{\prime}$ & \cite{1971He08} \\
$^{56}$Fe & 0.118(12) &  &  & $^{56}$Fe & $^{16}$O/$^{35}$Cl & 21-30; 60-79  & CE* & \cite{1971DaZM} \\
$^{56}$Fe & 0.125(27) &  &  & $^{56}$Fe & e- & 60.2 & EE$^{\prime}$ & \cite{1970Pe15} \\
$^{56}$Fe &  & 10.3(20) &  & $^{56}$Fe & $^{16}$O & 14-35  & TDSA & \cite{1969Sp05} \\
$^{56}$Fe &  & 11.3$^{+40}_{-24}$ &  & $^{56}$Fe & $^{16}$O & 34 & TDSA & \cite{1965Es01} \\
$^{56}$Fe & 0.097(10) &  &  & $^{56}$Fe & $^{16}$O & 33 & CE* & \cite{1964El03} \\
$^{56}$Fe &  & 8.5(29) &  & $^{56}$Fe & $\gamma$ & 0.5-3  & GG & \cite{1964Bo22} \\
$^{56}$Fe &  & 9.6(18) &  & $^{56}$Fe & $\gamma$ & 0.845-3.2  & GG & \cite{1963Be29} \\
$^{56}$Fe & 0.0720(35) &  &  & $^{56}$Fe & e- & 150 & EE$^{\prime}$ & \cite{1962Be18} \\
$^{56}$Fe &  & 10.6(17) &  & $^{56}$Fe & $\gamma$ &  & GG & \cite{1961Me11} \\
$^{56}$Fe &  & 8.6(29) &  & $^{56}$Fe & $\gamma$ &  & GG & \cite{1961Ke06} \\
$^{56}$Fe & 0.100(20) &  &  & $^{56}$Fe & $^{16}$O & 27 & CE & \cite{1960Go08} \\
$^{56}$Fe & 0.061(12) &  &  & $^{56}$Fe & $^{14}$N & 16.3, 36  & CE* & \cite{1960An07} \\
$^{56}$Fe & 0.100(25) &  &  & $^{56}$Fe & $^{16}$O & 39 & CE* & \cite{1960Ad01} \\
$^{56}$Fe & 0.070(18) &  &  & $^{56}$Fe & N & 15.9-35  & CE* & \cite{1959Al95} \\
$^{56}$Fe & 0.100(20) &  &  & $^{56}$Fe & $^{4}$He & 6 & CE & \cite{1956Te26} \\
$^{58}$Fe & 0.1234(36) &  &  & $^{58}$Fe & $^{12}$C/$^{52}$Cr & 22, 110-120  & CE* & \cite{1981Le02} \\
$^{58}$Fe &  & 3.4$^{+10}_{-9}$ &  & $^{58}$Fe & $^{4}$He & 10 & TDSA & \cite{1978Bo35} \\
$^{58}$Fe & 0.086(5) &  &  & $^{58}$Fe & $^{40}$Ca & 76 & CE* & \cite{1974ToZJ} \\
$^{58}$Fe & 0.094(8) &  &  & $^{58}$Fe & e- & 150, 225  & EE$^{\prime}$ & \cite{1972Li28} \\
$^{58}$Fe & 0.110(22) &  &  & $^{58}$Fe & $^{14}$N & 16.3 & CE & \cite{1960An07} \\
$^{58}$Fe & 0.20(5) &  &  & $^{58}$Fe & N & 15.9-35  & CE* & \cite{1959Al95} \\
$^{60}$Fe &  & 11.4(12) &  & $^{64}$Ni & $^{238}$U & 6.5 A & RDM & \cite{2010Lj01} \\
$^{60}$Fe &  & 11.6(22) &  & $^{48}$Ca & $^{15}$N/$^{18}$O & 25-55  & RDM & \cite{1977Wa10} \\
$^{62}$Fe &  & 8.0(10) &  & $^{197}$Au & $^{62}$Fe & 97.8 A & RDM & \cite{2011Ro02} \\
$^{62}$Fe &  & 7.4(9) &  & $^{64}$Ni & $^{238}$U & 6.5 A & RDM & \cite{2010Lj01} \\
$^{64}$Fe &  & 10.3(10) &  & $^{197}$Au & $^{64}$Fe & 95 A & RDM & \cite{2011Ro02} \\
$^{64}$Fe &  & 7.4(26) &  & $^{64}$Ni & $^{238}$U & 6.5 A & RDM & \cite{2010Lj01} \\
$^{66}$Fe &  & 39.4(40) &  & $^{197}$Au & $^{66}$Fe & 88.3 A & RDM & \cite{2011Ro02} \\
$^{54}$Ni & 0.059(17) &  &  & Pb & $^{54}$Ni & 42 A & CE* & \cite{2005Ya26} \\
$^{54}$Ni & 0.063(17) &  &  & $^{197}$Au & $^{54}$Ni & 70.3 A & CE* & \cite{2004Yu10} \\
$^{56}$Ni & 0.049(12) &  &  & $^{197}$Au & $^{56}$Ni & 85.8 A & CE* & \cite{2004Yu10} \\
$^{56}$Ni &  &  & 0.144(34) & $^{208}$Pb & $^{56}$Ni & 70.7 A & CE* & \cite{1998YaZR} \\
$^{56}$Ni & 0.060(12) &  &  & $^{1}$H & $^{56}$Ni & 101 A & IN-EL & \cite{1995Kr17} \\
$^{56}$Ni &  & 0.076$^{+49}_{-24}$ &  & $^{54}$Fe & $^{3}$He & 10 & TDSA & \cite{1973Sc28} \\
$^{58}$Ni & 0.0662(50) &  &  & $^{58}$Ni & $^{6}$Li & 240 & IN-EL & \cite{2010Kr01} \\
$^{58}$Ni & 0.0728(50) &  &  & $^{58}$Ni & $^{6}$Li & 240 & IN-EL & \cite{2010Kr01} \\
$^{58}$Ni &  & 1.00$^{+15}_{-10}$ &  & Ni & n & 1.6,1.8  & TDSA & \cite{2008Or02} \\
$^{58}$Ni & 0.0707(145) &  &  & $^{197}$Au & $^{58}$Ni & 77.8 A & CE* & \cite{2004Yu10} \\
$^{58}$Ni &  & 1.27(2) &  & $^{12}$C & $^{58}$Ni & 155, 160  & TDSA & \cite{2001Ke08} \\
$^{58}$Ni & 0.0588(40) &  &  & $^{58}$Ni & e- & 124, 180  & EE$^{\prime}$ & \cite{1983Kl09} \\
$^{58}$Ni &  & 0.90(11) &  & $^{58}$Ni & $\gamma$ & 0.5-1.65  & GG & \cite{1981Ca10} \\
$^{58}$Ni &  & 0.92(17) &  & $^{58}$Ni & p & 8 & TDSA & \cite{1973BeYD} \\
$^{58}$Ni & 0.0660(40) &  &  & $^{58}$Ni & $^{16}$O & 35-60  & CE* & \cite{1973Ch13} \\
$^{58}$Ni &  & 1.07(8) &  & $^{58}$Ni & $\gamma$ &  & GG & \cite{1972ArZD} \\
$^{58}$Ni & 0.0680(20) &  &  & $^{58}$Ni & $^{16}$O & 30,32,34  & CE* & \cite{1971ChZF} \\
$^{58}$Ni &  & 0.98(9) &  & $^{58}$Ni & $\gamma$ & $<$4.5  & GG & \cite{1970Me18} \\
$^{58}$Ni & 0.0731(17) &  &  & $^{58}$Ni & $^{12}$C/$^{16}$O/$^{32}$S & 21-22, 25-30, 60-70  & CE* & \cite{1970Le17} \\
$^{58}$Ni & 0.0554(30) &  &  & $^{58}$Ni & e- & 150, 225  & EE$^{\prime}$ & \cite{1969Af01} \\
$^{58}$Ni &  & 0.94(12) &  & $^{58}$Ni & p & 7-9.0 & TDSA & \cite{1969Be48} \\
$^{58}$Ni & 0.0657(11) &  &  & $^{58}$Ni & e- & 45-65  & EE$^{\prime}$ & \cite{1967Du07} \\
$^{58}$Ni &  & 0.62(20) &  & $^{58}$Ni & $\gamma$ & 0.5-3.0  & GG & \cite{1964Bo22} \\
$^{58}$Ni & 0.072(7) &  &  & $^{58}$Ni & $^{4}$He & 4.5-8  & CE & \cite{1962St02} \\
$^{58}$Ni & 0.098(13) &  &  & $^{58}$Ni & e- & 183 & EE$^{\prime}$ & \cite{1961Cr01} \\
$^{58}$Ni & 0.063(13) &  &  & $^{58}$Ni & $^{16}$O & 34 & CE* & \cite{1960Go08} \\
$^{58}$Ni & 0.080(16) &  &  & $^{58}$Ni & $^{14}$N & 36 & CE* & \cite{1960An07} \\
$^{58}$Ni & 0.071(14) &  &  & $^{58}$Ni & $^{4}$He &  & CE? & \cite{1960An07} \\
$^{58}$Ni & 0.100(25)$^{s}$ &  &  & $^{58}$Ni & N & 15.9-35  & CE & \cite{1959Al95} \\
$^{60}$Ni &  & 1.30$^{+30}_{-20}$ &  & $^{60}$Ni & n & 1.6,1.8  & TDSA & \cite{2008Or02} \\
$^{60}$Ni &  & 1.31(3) &  & $^{12}$C & $^{60}$Ni & 155, 160  & TDSA & \cite{2001Ke02} \\
$^{60}$Ni &  & 1.30(36)  &  & N/A & N/A & N/A & TCS & \cite{1976Kl04} \\
$^{60}$Ni & 0.1020(40) &  &  & $^{60}$Ni & e- & 30-60  & EE$^{\prime}$ & \cite{1974Ye01} \\
$^{60}$Ni & 0.087(7) &  &  & $^{60}$Ni & e- & 45-250  & EE$^{\prime}$ & \cite{1974Si01} \\
$^{60}$Ni &  & 1.00(7) &  & $^{60}$Ni & $^{35}$Cl & 56-68  & TDSA & \cite{1973Fi15} \\
$^{60}$Ni & 0.082(6) &  &  & $^{60}$Ni & $\gamma$ &  & GG & \cite{1972ArZD} \\
$^{60}$Ni & 0.0910(30) &  &  & $^{60}$Ni & $^{16}$O & 30,32,34  & CE & \cite{1971ChZF} \\
$^{60}$Ni & 0.092(12) &  &  & $^{60}$Ni & $\gamma$ & $<$4.5  & GG & \cite{1970Me18} \\
$^{60}$Ni & 0.0938(20) &  &  & $^{60}$Ni & $\gamma$ & 1.333 & GG & \cite{1970Me08} \\
$^{60}$Ni & 0.0914(20) &  &  & $^{60}$Ni & $^{16}$O/$^{32}$S & 28-70  & CE* & \cite{1969Cl05} \\
$^{60}$Ni & 0.0603(28) &  &  & $^{60}$Ni & e- & 150, 225  & EE$^{\prime}$ & \cite{1969Af01} \\
$^{60}$Ni & 0.077(8) &  &  & $^{60}$Ni & e- & 183,250 & EE$^{\prime}$ & \cite{1969To08} \\
$^{60}$Ni & 0.108(21) &  &  & $^{60}$Ni & $\gamma$ & 1-2.0 & GG & \cite{1967Be39} \\
$^{60}$Ni & 0.0845(9) &  &  & $^{60}$Ni & e- & 45-65  & EE$^{\prime}$ & \cite{1967Du07} \\
$^{60}$Ni & 0.091(5) &  &  & $^{60}$Ni & $^{4}$He & 4.5-8  & CE & \cite{1962St02} \\
$^{60}$Ni & 0.123(15) &  &  & $^{60}$Ni & e- & 183 & EE$^{\prime}$ & \cite{1961Cr01} \\
$^{60}$Ni & 0.11(1) &  &  & $^{60}$Ni & $^{14}$N & 36 & CE* & \cite{1960An07} \\
$^{60}$Ni & 0.120(24) &  &  & $^{60}$Ni & $^{16}$O & 34 & CE* & \cite{1960Go08} \\
$^{60}$Ni &  & 1.0(3) &  & $^{60}$Ni & $\gamma$ & 133 & GG & \cite{1959Bu12} \\
$^{60}$Ni & 0.160(40) &  &  & $^{60}$Ni & $^{14}$N & 15.9-35  & CE & \cite{1959Al95} \\
$^{60}$Ni &  & 1.1(2) &  & $^{60}$Ni & $\gamma$ & 1.33,1.17  & GG & \cite{1956Me59} \\
$^{62}$Ni &  & 1.79$^{+86}_{-48}$ &  & $^{62}$Ni & n & 2.8-4.1  & TDSA & \cite{2011Ch05} \\
$^{62}$Ni &  & 2.01(7) &  & $^{12}$C & $^{62}$Ni & 155, 160  & TDSA & \cite{2001Ke02} \\
$^{62}$Ni &  & 2.15(42) &  & $^{62}$Ni & $\gamma$ & 0.5-1.65  & GG & \cite{1981Ca10} \\
$^{62}$Ni &  & 1.55(25) &  & $^{59}$Co & $^{4}$He & 10 & TDSA & \cite{1978Ke11} \\
$^{62}$Ni &  & 1.55(25)$^{d}$ &  & $^{59}$Co & $^{4}$He & 10 & TDSA & \cite{1978KlZR} \\
$^{62}$Ni &  & 1.7(5) &  & $^{59}$Co & $^{4}$He & 8 & TDSA & \cite{1977OhZX} \\
$^{62}$Ni &  & 2.1(5)$^{d}$ &  & $^{62}$Ni & $\gamma$ & 0.5-1.65  & GG & \cite{1977Ca14} \\
$^{62}$Ni & 0.102(10) &  &  & $^{62}$Ni & e- & 50,60,80,90  & EE$^{\prime}$ & \cite{1975DeXW} \\
$^{62}$Ni & 0.0618(42) &  &  & $^{62}$Ni & e- & 150225 & EE$^{\prime}$ & \cite{1972Li28} \\
$^{62}$Ni & 0.0880(30) &  &  & $^{62}$Ni & $^{16}$O & 30, 32, 34  & CE & \cite{1971ChZF} \\
$^{62}$Ni & 0.0899(28) &  &  & $^{62}$Ni & $^{12}$C/$^{16}$O/$^{32}$S & 70 & CE* & \cite{1970Le17} \\
$^{62}$Ni & 0.084(5) &  &  & $^{62}$Ni & $^{28}$Si & 70 & CE* & \cite{1969Ha31} \\
$^{62}$Ni & 0.0877(11) &  &  & $^{62}$Ni & e- & 45-65  & EE$^{\prime}$ & \cite{1967Du07} \\
$^{62}$Ni &  & 2.28(18) &  & $^{62}$Ni & $^{16}$O & 36 & TDSA & \cite{1965Es01} \\
$^{62}$Ni & 0.083(8) &  &  & $^{62}$Ni & $^{4}$He & 4-8.0 & CE & \cite{1962St02} \\
$^{62}$Ni & 0.085(17) &  &  & $^{62}$Ni & $^{14}$N & 36 & CE* & \cite{1960An07} \\
$^{62}$Ni & 0.140(35) &  &  & $^{62}$Ni & $^{14}$N & 15.9-35  & CE & \cite{1959Al95} \\
$^{64}$Ni &  & 1.57(5) &  & $^{12}$C & $^{64}$Ni & 155, 160  & TDSA & \cite{2001Ke08} \\
$^{64}$Ni &  & 0.025(12) &  & $^{64}$Ni & n & fast  & TDSA$^{r}$ & \cite{1989Ge09} \\
$^{64}$Ni & 0.0744(20) &  &  & $^{64}$Ni & e- & 147.4-356  & EE$^{\prime}$ & \cite{1988Br10} \\
$^{64}$Ni &  & 0.40(15) &  & $^{64}$Ni & $^{4}$He & 13 & TDSA$^{r}$ & \cite{1974Iv01} \\
$^{64}$Ni & 0.0650(40) &  &  & $^{64}$Ni & $^{16}$O & 30,32,34  & CE & \cite{1971ChZF} \\
$^{64}$Ni & 0.0650(34) &  &  & $^{64}$Ni & e- & 150, 225  & EE$^{\prime}$ & \cite{1969Af01} \\
$^{64}$Ni & 0.087(17) &  &  & $^{64}$Ni & $^{14}$N & 36 & CE* & \cite{1960An07} \\
$^{64}$Ni & 0.077(15) &  &  & $^{64}$Ni & $^{4}$He &  & CE? & \cite{1960An07} \\
$^{64}$Ni & 0.090(18)$^{s}$ &  &  & $^{64}$Ni & N & 15.9-35  & CE & \cite{1959Al95} \\
$^{66}$Ni & 0.06(1) &  &  & $^{208}$Pb & $^{66}$Ni & 0.3 c & CE* & \cite{2002So03} \\
$^{66}$Ni & 0.09(1) &  &  & $^{58}$Ni & $^{70}$Ni & 65.9 A & CE* & \cite{2000LeZZ} \\
$^{68}$Ni & 0.028(11) &  &  & $^{108}$Pd & $^{68}$Ni & 2.9 A & CE* & \cite{2008Br18} \\
$^{68}$Ni & 0.0255(60) &  &  & $^{208}$Pb & $^{66}$Ni & 0.3 c & CE* & \cite{2002So03} \\
$^{68}$Ni & 0.029(7) &  &  & $^{58}$Ni & $^{70}$Ni & 65.9 A & CE* & \cite{2000LeZZ} \\
$^{70}$Ni & 0.086(14) &  &  & $^{208}$Pb & $^{70}$Ni & 0.28 c & CE* & \cite{2006Pe13} \\
$^{74}$Ni &  &  & 0.21(3) & p & $^{74}$Ni & 81 A & IN-EL & \cite{2010Ao01} \\
$^{62}$Zn &  & 4.2(7) &  & $^{63}$Zn & C & 0.35 c & RDM & \cite{2007St16} \\
$^{62}$Zn &  & 4.3(3) &  & $^{62}$Zn & Fe & 160 & TDSA & \cite{2002Ke02} \\
$^{62}$Zn &  & 4.20(30) &  & $^{6}$Li & $^{58}$Ni & 15-24  & TDSA & \cite{1981Wa09} \\
$^{62}$Zn &  & 2.5$^{+10}_{-20}$ &  & $^{4}$He & $^{61}$Ni & 30 & TDSA & \cite{1977BrYO} \\
$^{64}$Zn &  & 2.85(9) &  & $^{64}$Zn & C & 180 & TDSA & \cite{2005Le12} \\
$^{64}$Zn &  & 2.70(8) &  & $^{64}$Zn & Fe, C & 160 & TDSA & \cite{2002Ke02} \\
$^{64}$Zn & 0.112(6) &  &  & p & $^{64}$Zn & 2-4.5  & CE* & \cite{1998Si25} \\
$^{64}$Zn & 0.168(4) &  &  & $^{4}$He/$^{16}$O/$^{18}$O & $^{64}$Zn & 8,35,30  & CE & \cite{1988Sa32} \\
$^{64}$Zn &  & 2.97(25) &  & $\gamma$ & $^{64}$Zn & 1.65 & GG & \cite{1981Ca10} \\
$^{64}$Zn &  & 3.00(30)$^{s}$ &  & $\gamma$ & $^{64}$Zn & 1.65 & GG & \cite{1977Ca14} \\
$^{64}$Zn &  & 4.0(10) &  & $^{16}$O & $^{51}$V & 49 & RDM & \cite{1977Al14} \\
$^{64}$Zn & 0.162(9) &  &  & e- & $^{64}$Zn & 100-275  & EE$^{\prime}$ & \cite{1977Ne05} \\
$^{64}$Zn & 0.155(9) &  &  & e- & $^{64}$Zn & 40-112  & EE$^{\prime}$ & \cite{1976Ne06} \\
$^{64}$Zn &  & 2.9(7)  &  & $^{4}$He & $^{61}$Ni & 6.4,8  & TDSA & \cite{1976Ch11} \\
$^{64}$Zn & 0.161(12) &  &  & $^{4}$He & $^{64}$Zn & 3-5.0 & CE & \cite{1975Th01} \\
$^{64}$Zn & 0.176(21)  &  &  & $^{35}$Cl & $^{64}$Zn & 56-68  & CE & \cite{1973Fi15} \\
$^{64}$Zn & 0.155(11)  &  &  & $\gamma$ & $^{64}$Zn &  & GG & \cite{1972ArZD} \\
$^{64}$Zn & 0.170(16) &  &  & e- & $^{64}$Zn & 150, 225  & EE$^{\prime}$ & \cite{1970Af04} \\
$^{64}$Zn & 0.108(5) &  &  & $\gamma$ & $^{64}$Zn &  & GG & \cite{1965Ta13} \\
$^{64}$Zn & 0.162(10) &  &  & $^{4}$He & $^{64}$Zn & 4.5-8  & CE & \cite{1962St02} \\
$^{64}$Zn & 0.110(22) &  &  & $^{14}$N & $^{64}$Zn & 36 & CE* & \cite{1960An07} \\
$^{64}$Zn & 0.110(22) &  &  & $^{4}$He & $^{64}$Zn & 6-7.0 & CE & \cite{1956Te26} \\
$^{66}$Zn &  & 2.5(1) &  & $^{66}$Zn & C & 180 & TDSA & \cite{2006Le24} \\
$^{66}$Zn & 0.144(9) &  &  & $^{66}$Zn & Pb & 274.2 & CE* & \cite{2003Ko51} \\
$^{66}$Zn &  & 2.43(5) &  & $^{66}$Zn & Fe, C & 160 & TDSA & \cite{2002Ke02} \\
$^{66}$Zn & 0.135(8) &  &  & p & $^{66}$Zn & 2-4.5  & CE* & \cite{1998Si25} \\
$^{66}$Zn &  & 2.71(23) &  & $\gamma$ & $^{66}$Zn & 1.65 & GG & \cite{1981Ca10} \\
$^{66}$Zn &  & 2.0(10) &  & $^{4}$He & $^{63}$Cu & 10, 16.7  & TDSA & \cite{1981Zh07} \\
$^{66}$Zn &  & 2.70(20)$^{s}$ &  & $\gamma$ & $^{66}$Zn & 1.65 & GG & \cite{1977Ca14} \\
$^{66}$Zn & 0.141(8) &  &  & e- & $^{66}$Zn & 100-275  & EE$^{\prime}$ & \cite{1977Ne05} \\
$^{66}$Zn &  & 2.5$^{+5}_{-2}$ &  & $^{4}$He & $^{64}$Ni & 27, 30  & TDSA & \cite{1977Mo20} \\
$^{66}$Zn & 0.137(10) &  &  & e- & $^{66}$Zn & 40-112  & EE$^{\prime}$ & \cite{1976Ne06} \\
$^{66}$Zn & 0.154(13) &  &  & $^{4}$He & $^{66}$Zn & 3-5.0 & CE & \cite{1975Th01} \\
$^{66}$Zn & 0.180(15) &  &  & e- & $^{66}$Zn & 225 & EE$^{\prime}$ & \cite{1973Li24} \\
$^{66}$Zn & 0.155(13) &  &  & $^{35}$Cl & $^{66}$Zn & 56-68  & CE & \cite{1973Fi15} \\
$^{66}$Zn &  & 2.2(9) &  & $^{4}$He & $^{66}$Zn & 25 & TDSA & \cite{1972Yo01} \\
$^{66}$Zn & 0.153(21) &  &  & $\gamma$ & $^{66}$Zn & 1.037 & GG & \cite{1972Ka22} \\
$^{66}$Zn & 0.138(16) &  &  & $\gamma$ & $^{66}$Zn &  & GG & \cite{1972ArZD} \\
$^{66}$Zn & 0.145(15) &  &  & e- & $^{66}$Zn & 150, 225  & EE$^{\prime}$ & \cite{1970Af04} \\
$^{66}$Zn & 0.15(6) &  &  & $\gamma$ & $^{66}$Zn & 1-2.0 & GG & \cite{1967Be39} \\
$^{66}$Zn & 0.145(13) &  &  & $^{4}$He & $^{66}$Zn & 4-8.0 & CE & \cite{1962St02} \\
$^{66}$Zn & 0.110(22) &  &  & $^{14}$N & $^{66}$Zn & 36 & CE* & \cite{1960An07} \\
$^{66}$Zn & 0.087(17) &  &  & $^{4}$He & $^{66}$Zn & 6-7.0 & CE & \cite{1956Te26} \\
$^{68}$Zn &  & 2.32(5) &  & $^{68}$Zn & C & 180 & TDSA & \cite{2005Le12} \\
$^{68}$Zn & 0.129(8) &  &  & $^{68}$Zn & Pb & 276 & CE* & \cite{2004Ko03} \\
$^{68}$Zn &  & 2.32(7) &  & $^{68}$Zn & Fe, C & 160 & TDSA & \cite{2002Ke02} \\
$^{68}$Zn & 0.105(7) &  &  & p & $^{68}$Zn & 2-4.5  & CE* & \cite{1998Si25} \\
$^{68}$Zn &  & 2.71(23)$^{s}$ &  & $\gamma$ & $^{68}$Zn & 1.65 & GG & \cite{1981Ca10} \\
$^{68}$Zn & 0.125(11) &  &  & e- & $^{68}$Zn & 100-275  & EE$^{\prime}$ & \cite{1977Ne05} \\
$^{68}$Zn & 0.105(8) &  &  & $\gamma$ & $^{68}$Zn & 1.65 & GG & \cite{1977Ca14} \\
$^{68}$Zn & 0.111(8) &  &  & e- & $^{68}$Zn & 40-112  & EE$^{\prime}$ & \cite{1976Ne06} \\
$^{68}$Zn &  & 1.3(3) &  & $^{4}$He & $^{68}$Zn & 13 & TDSA & \cite{1974Iv01} \\
$^{68}$Zn & 0.126(13) &  &  & $^{35}$Cl & $^{68}$Zn & 56-68  & CE & \cite{1973Fi15} \\
$^{68}$Zn & 0.108(14) &  &  & e- & $^{68}$Zn & 225 & EE$^{\prime}$ & \cite{1973Li24} \\
$^{68}$Zn & 0.140(16) &  &  & $\gamma$ & $^{68}$Zn &  & GG & \cite{1972ArZD} \\
$^{68}$Zn & 0.125(11) &  &  & $^{4}$He & $^{68}$Zn & 4-8.0 & CE & \cite{1962St02} \\
$^{68}$Zn & 0.110(22) &  &  & $^{14}$N & $^{68}$Zn & 36 & CE* & \cite{1960An07} \\
$^{70}$Zn & 0.164(28) &  &  & $^{70}$Zn & $^{58}$Ni & 65.9 A & CE* & \cite{2002So03} \\
$^{70}$Zn &   & 5.3(3) &  & $^{70}$Zn & Fe, C & 160 & TDSA & \cite{2002Ke02} \\
$^{70}$Zn & 0.235(25) &  &  & p & $^{70}$Zn & 2-4.5  & CE* & \cite{1998Si25} \\
$^{70}$Zn & 0.205(19) &  &  & e- & $^{70}$Zn & 40-112  & EE$^{\prime}$ & \cite{1976Ne06} \\
$^{70}$Zn & 0.160(14) &  &  & $^{4}$He & $^{70}$Zn & 4-8.0 & CE & \cite{1962St02} \\
$^{72}$Zn & 0.174(21) &  &  & $^{72}$Zn & Pb & 35 A & CE* & \cite{2002Le17} \\
$^{74}$Zn &  & 27.6(43) &  & $^{9}$Be & $^{76}$Ge & 60 A & RDM & \cite{2011Ni03} \\
$^{74}$Zn & 0.201(16) &  &  & $^{74}$Zn & $^{108}$Pd,$^{120}$Sn & 2.87 A & CE* & \cite{2007Va20} \\
$^{74}$Zn & 0.204(15) &  &  & $^{74}$Zn & $^{208}$Pb & 0.28 c & CE* & \cite{2006Pe13} \\
$^{76}$Zn & 0.145(18) &  &  & $^{76}$Zn & $^{108}$Pd,$^{120}$Sn & 2.83 A & CE* & \cite{2007Va20} \\
$^{78}$Zn & 0.077(19) &  &  & $^{78}$Zn & $^{108}$Pd,$^{120}$Sn & 2.87 A & CE* & \cite{2007Va20} \\
$^{80}$Zn & 0.073(9) &  &  & $^{80}$Zn & $^{108}$Pd,$^{120}$Sn & 2.79 A & CE* & \cite{2009Va01,2007Va20} \\

\hline \hline
\end{longtable}
\newpage

\begin{theDTbibliography}{2005Ya26}
\bibitem[2011Ch05]{2011Ch05} A. Chakraborty, J.N. Orce, S.F. Ashley  {\it et al.}, 
\newblock Phys.Rev. {\bf C 83}, 034316 (2011).
\bibitem[2011Ni03]{2011Ni03} M. Niikura, B. Mouginot, F. Azaiez  {\it et al.}, 
\newblock Acta Phys. Pol. {\bf B 42}, 537 (2011).
\bibitem[2011Ro02]{2011Ro02} W. Rother, A. Dewald, H. Iwasaki {\it et al.}, 
\newblock Phys. Rev. Lett {\bf 106}, 022502 (2011).
\bibitem[2010Lj01]{2010Lj01} J. Ljungvall, A. Gorgen, A. Obertelli {\it et al.}, 
\newblock Phys.Rev. {\bf C 81}, 061301 (2010).
\bibitem[2010Kr01]{2010Kr01} Krishichayan, X. Chen, Y.-W. Lui, Y. Tokimoto {\it et al.}, 
\newblock Phys. Rev. {\bf C 81}, 014603 (2010).
\bibitem[2010Ao01]{2010Ao01} N. Aoi, S. Kanno, S. Takeuchi {\it et al.}, 
\newblock Phys. Lett. {\bf B 692} 302 (2010).
\bibitem[2009Ao01]{2009Ao01}  N. Aoi, E. Takeshita, H. Suzuki {\it et al.}, 
\newblock Phys.Rev.Lett. {\bf 102}, 012502 (2009).
\bibitem[2009Va01]{2009Va01} J. Van de Walle, F. Aksouh, T. Behrens {\it et al.}, 
\newblock Phys.Rev. {\bf C 79}, 014309 (2009).
\bibitem[2008Or02]{2008Or02} J.N. Orce, B. Crider, S. Mukhopadhyay {\it et al.}, 
\newblock Phys. Rev. {\bf C 77}, 064301 (2008).
\bibitem[2008Br18]{2008Br18} N. Bree, I. Stefanescu, P.A. Butler {\it et al.}, 
\newblock Phys. Rev. {\bf C 78}, 047301 (2008).
\bibitem[2007St16]{2007St16} K. Starosta, A. Dewald, A. Dunomes {\it et al.}, 
\newblock Phys.Rev.Lett. {\bf 99}, 042503 (2007).
\bibitem[2007Va20]{2007Va20} J. Van de Walle, F. Aksouh, F. Ames {\it et al.}, 
\newblock Phys.Rev.Lett. {\bf 99}, 142501 (2007).
\bibitem[2006Pe13]{2006Pe13} O. Perru, O. Sorlin, S. Franchoo {\it et al.}, 
\newblock Phys. Rev. Lett. {\bf 96}, 232501 (2006).
\bibitem[2006Le24]{2006Le24} J. Leske, K.-H. Speidel, S. Schielke {\it et al.}, 
\newblock Phys.Rev. {\bf C 73}, 064305 (2006).
\bibitem[2005Ya26]{2005Ya26} K. Yamada, T. Motobayashi, N. Aoi {\it et al.}, 
\newblock Eur.Phys.J. {\bf A 25}, Supplement 1, 409 (2005).
\bibitem[2005Bu29]{2005Bu29}  A. Burger, T.R. Saito, H. Grawe {\it et al.}, 
\newblock Phys.Lett. {\bf B 622}, 29 (2005).
\bibitem[2005Le12]{2005Le12} J. Leske, K.-H. Speidel, S. Schielke {\it et al.}, 
\newblock Phys.Rev. {\bf C 71}, 034303 (2005).
\bibitem[2004Yu07]{2004Yu07}  K.L. Yurkewicz, D. Bazin, B.A. Brown {\it et al.}, 
\newblock Phys.Rev. {\bf C 70}, 034301 (2004).
\bibitem[2004Yu10]{2004Yu10} K.L. Yurkewicz, D. Bazin, B.A. Brown {\it et al.}, 
\newblock Phys. Rev. {\bf C 70}, 054319 (2004). 
\bibitem[2004Ko03]{2004Ko03} M. Koizumi, A. Seki, Y. Toh {\it et al.}, 
\newblock Nucl.Phys. {\bf A730}, 46 (2004).
\bibitem[2003Ko51]{2003Ko51} M.Koizumi, A.Seki, Y.Toh {\it et al.}, 
\newblock Eur.Phys.J. {\bf A 18}, 87 (2003).
\bibitem[2002So03]{2002So03} O. Sorlin, S. Leenhardt, C. Donzaud {\it et al.}, 
\newblock Phys.Rev. Lett. {\bf 88}, 092501 (2002).
\bibitem[2002Ke02]{2002Ke02} O. Kenn, K.-H. Speidel, R. Ernst {\it et al.}, 
\newblock Phys.Rev. {\bf C65}, 034308 (2002).
\bibitem[2002Le17]{2002Le17} S. Leenhardt, O. Sorlin, M.G. Porquet {\it et al.}, 
\newblock Eur.Phys.J. {\bf A 14}, 1 (2002).
\bibitem[2001Ke02]{2001Ke02} O. Kenn, K.-H. Speidel, R. Ernst {\it et al.}, 
\newblock Phys. Rev. {\bf C 63}, 021302 (2001).
\bibitem[2001Ke08]{2001Ke08} O. Kenn, K.-H. Speidel, R. Ernst {\it et al.}, 
\newblock Phys. Rev. {\bf C 63}, 064306 (2001).
\bibitem[2000Er01]{2000Er01} R. Ernst, K.-H. Speidel, O. Kenn {\it et al.}, 
\newblock Phys.Rev.Lett. {\bf 84}, 416 (2000).
\bibitem[2000LeZZ]{2000LeZZ} S. Leenhardt, C. Donzaud, F. Amorini, {\it et al.}, 
\newblock Univ.Paris, Inst.Phys.Nucl., 1998-1999 Ann.Rept., p.29 (2000).
\bibitem[1998YaZR]{1998YaZR} Y. Yanagisawa, T. Motobayashi, S. Shimoura {\it et al.}, 
\newblock Proc. Conf on Exotic Nuclei and Atomic Masses, Bellaire, Michigan, June 23-27, 1998, p.610 (1998); AIP Conf. Proc. {\bf 455} (1998). 
\bibitem[1998Si25]{1998Si25} K.P. Singh, D.C. Tayal, H.S. Hans, 
\newblock Phys.Rev. {\bf C58}, 1980 (1998).
\bibitem[1996Ch03]{1996Ch03} L.C. Chamon, D. Pereira, E.S. Rossi {\it et al.},
\newblock Nucl. Phys. {\bf A 597}, 253 (1996).
\bibitem[1995Kr17]{1995Kr17} G. Kraus, P. Egelhof, C. Fischer {\it et al.}, 
\newblock Phys. Scr. {\bf T 56}, 114 (1995).
\bibitem[1989Va02]{1989Va02} P.J. van Hall, S.D. Wassenaar, S.S. Klein {\it et al.},
\newblock J. Phys. (London) {\bf G 15}, 199 (1989).
\bibitem[1989Ge09]{1989Ge09} M.K. Georgieva, D.V. Elenkov, D.P. Lefterov, G.H. Toumbev,
\newblock Fiz.Elem.Chastits At.Yadra 20, 930 (1989); Sov.J.Part.Nucl. 20, 393 (1989).
\bibitem[1988Br10]{1988Br10} M.R. Braunstein, J.J. Kraushaar, R.P. Michel {\it et al.}, 
\newblock Phys. Rev. {\bf C 37}, 1870 (1988).
\bibitem[1988Sa32]{1988Sa32} S.Salem-Vasconcelos, M.J.Bechara, J.H.Hirata, O.Dietzsch, 
\newblock Phys.Rev. {\bf C38}, 2439 (1988).
\bibitem[1983Li02]{1983Li02}  J.W. Lightbody, Jr., J.W. Lightbody, J.B. Bellicard {\it et al.}, 
\newblock Phys.Rev. {\bf C27}, 113 (1983).
\bibitem[1983Kl09]{1983Kl09} R. Klein, P. Grabmayr, Y. Kawazoe {\it et al.}, 
\newblock Nuovo Cim. {\bf 76 A}, 369 (1983).
\bibitem[1981Ah02]{1981Ah02} J. Ahlert, M. Schumacher, 
\newblock Z.Phys. {\bf A301}, 75 (1981).
\bibitem[1981Le02]{1981Le02}  M.J. LeVine, E.K. Warburton, D. Schwalm, 
\newblock Phys.Rev. {\bf C23}, 244 (1981).
\bibitem[1981Ca10]{1981Ca10} Y. Cauchois, H. Ben Abdelaziz, R. Kherouf, C. Schloesing-Moller, 
\newblock J. Phys.(London) {\bf G 7}, 1539 (1981).
\bibitem[1981Wa09]{1981Wa09} N.J. Ward, L.P. Ekstrom, G.D. Jones {\it et al.}, 
\newblock J.Phys.(London) {\bf G7}, 815 (1981).
\bibitem[1981Zh07]{1981Zh07} U.Yu. Zhovliev, M.F. Kudoyarov, I.Kh. Lemberg, A.A. Pasternak, 
\newblock Izv.Akad.Nauk SSSR, Ser.Fiz. {\bf 45}, 1879 (1981).
\bibitem[1979Ek03]{1979Ek03} L.P. Ekstrom, G.D. Jones, F. Kearns {\it et al.}, 
\newblock J.Phys.(London) {\bf G5}, 803 (1979).
\bibitem[1978Po04]{1978Po04}  V.N. Polishchuk, N.G. Shevchenko, N.G. Afanasev {\it et al.}, 
\newblock Yad.Fiz. {\bf 27}, 1145 (1978); Sov.J.Nucl.Phys. {\bf 27}, 607 (1978).
\bibitem[1978Bo35]{1978Bo35} H.H. Bolotin, A.E. Stuchbery, K. Amos, I. Morrison, 
\newblock Nucl.Phys. {\bf A311}, 75 (1978).
\bibitem[1978Ke11]{1978Ke11} D.L. Kennedy, H.H. Bolotin, I. Morrison, K. Amos, 
\newblock Nucl. Phys. {\bf A 308}, 14 (1978).
\bibitem[1978KlZR]{1978KlZR} D.L. Kennedy, H.H. Bolotin, I. Morrison, K. Amos, 
\newblock UM-P-88, p.9 (1978).
\bibitem[1977Wa10]{1977Wa10} E.K.Warburton, J.W.Olness, A.M.Nathan {\it et al.}, 
\newblock Phys.Rev. {\bf C16}, 1027 (1977).
\bibitem[1977OhZX]{1977OhZX} H. Ohnuma, J. Kasagi, Y. Iritani {\it et al.}, 
\newblock Proc. Int. Conf. Nucl. Structure, Tokyo, Japan, Int. Academic Printing Co., Ltd. Japan, Vol.1, 270 (1977).
\bibitem[1977Ca14]{1977Ca14} Y. Cauchois, H. ben Abdelaziz, Y. Heno {\it et al.}, 
\newblock C.R. Acad. Sci., Ser. {\bf B 284}, 65 (1977).
\bibitem[1977BrYO]{1977BrYO} J.F. Bruandet, Tsan Ung Chan, C. Morand {\it et al.}, 
\newblock Int.Symp.High-Spin States, Nucl.Struct., Dresden, L.Funke, Ed., ZfK-336, p.119 (1977).
\bibitem[1977Al14]{1977Al14} A.A.Aleksandrov, V.S.Zvonov, M.A.Ivanov {\it et al.}, 
\newblock Izv.Akad.Nauk SSSR, Ser.Fiz. {\bf 41}, 49 (1977); Bull.Acad.Sci.USSR, Phys.Ser. {\bf 41}, No.1, 39 (1977).
\bibitem[1977Ne05]{1977Ne05} R. Neuhausen, 
\newblock Nucl.Phys. {\bf A282}, 125 (1977).
\bibitem[1977Mo20]{1977Mo20} C. Morand, J.F. Bruandet, A. Giorni, Tsan Ung Chan, 
\newblock J.Phys.(Paris) {\bf 38}, 1319 (1977).
\bibitem[1976Li19]{1976Li19}  J.W. Lightbody, Jr., J.W. Lightbody, S. Penner {\it et al.}, 
\newblock Phys.Rev. {\bf C14}, 952 (1976).
\bibitem[1976Kl04]{1976Kl04} A. Kluge, W. Thomas, 
\newblock Nucl. Instrum. Methods {\bf 134}, 525 (1976).
\bibitem[1976Ne06]{1976Ne06} R. Neuhausen, J.W. Lightbody, Jr., J.W. Lightbody {\it et al.}, 
\newblock Nucl.Phys. {\bf A263}, 249 (1976).
\bibitem[1976Ch11]{1976Ch11} A. Charvet, R. Duffait, T. Negadi {\it et al.}, 
\newblock Phys.Rev. {\bf C13}, 2237 (1976).
\bibitem[1975Ha04]{1975Ha04} B. Haas, P. Taras, J.C. Merdinger, R. Vaillancourt, 
\newblock Nucl.Phys. {\bf A238}, 253 (1975).
\bibitem[1975To06]{1975To06} C.W. Towsley, D. Cline, R.N. Horoshko, 
\newblock Nucl.Phys. {\bf A250}, 381 (1975).
\bibitem[1975DeXW]{1975DeXW}  J.E.P. de Bie, C.W. de Jager, A.A.C. Klaasse {\it et al.}, 
\newblock IKO Progr.Rept.1975, p.3 (1975).
\bibitem[1975Th01]{1975Th01} M.J.Throop, Y.T.Cheng, D.K.McDaniels, 
\newblock Nucl.Phys. {\bf A239}, 333 (1975).
\bibitem[1974Br04]{1974Br04} B.A. Brown, D.B. Fossan, J.M. McDonald, K.A. Snover, 
\newblock Phys.Rev. {\bf C9}, 1033 (1974).
\bibitem[1974Po15]{1974Po15} A.R. Poletti, B.A. Brown, D.B. Fossan, E.K. Warburton, 
\newblock Phys.Rev. {\bf C10}, 2329 (1974).
\bibitem[1974ToZJ]{1974ToZJ} C.W. Towsley, 
\newblock Thesis, Univ.Rochester (1974); Diss.Abstr.Int. {\bf 35B}, 1864 (1974).
\bibitem[1974Ye01]{1974Ye01} R. Yen, L.S. Cardman, D. Kalinsky {\it et al.}, 
\newblock Nucl. Phys. {\bf A 235}, 135 (1974).
\bibitem[1974Si01]{1974Si01} R.P. Singhal, S.W. Brain, W.A. Gillespie {\it et al.}, 
\newblock Nucl. Phys. {\bf A 218}, 189 (1974).
\bibitem[1974Iv01]{1974Iv01} M. Ivascu, D. Popescu, E. Dragulescu {\it et al.}, 
\newblock Nucl. Phys. {\bf A 218}, 104 (1974).
\bibitem[1973Ku10]{1973Ku10} W. Kutschera, R.B. Huber, C. Signorini, P. Blasi, 
\newblock Nucl.Phys. {\bf A210}, 531 (1973).
\bibitem[1973De09]{1973De09} W. Dehnhardt, O.C. Kistner, W. Kutschera, H.J. Sann, 
\newblock Phys.Rev. {\bf C7}, 1471 (1973).
\bibitem[1973Sc28]{1973Sc28} N. Schulz, J. Chevallier, B. Haas {\it et al.}, 
\newblock Phys. Rev. {\bf C 8}, 1779 (1973).
\bibitem[1973BeYD]{1973BeYD} W. Beens, 
\newblock Thesis, Vrije Univ., Amsterdam (1973).
\bibitem[1973Ch13]{1973Ch13} P.R. Christensen, I. Chernov, E.E. Gross {\it et al.}, 
\newblock Nucl. Phys. {\bf A 207}, 433 (1973).
\bibitem[1973Fi15]{1973Fi15} T.R. Fisher, P.D. Bond, 
\newblock Part. Nucl. {\bf 6}, 119 (1973).
\bibitem[1973Li24]{1973Li24} A.S. Litvinenko, N.G. Shevchenko, N.G. Afanasev {\it et al.}, 
\newblock Yad.Fiz. {\bf 18}, 250 (1973); Sov.J.Nucl.Phys. {\bf 18}, 128 (1974).
\bibitem[1972Ra14]{1972Ra14}  S. Raman, R.L. Auble, W.T. Milner {\it et al.}, 
\newblock Nucl.Phys. {\bf A184}, 138 (1972).
\bibitem[1972WaYZ]{1972WaYZ} D. Ward, I.M. Szoghy, J.S. Forster, W.G. Davies, 
\newblock AECL-4314, p.9 (1972).
\bibitem[1972Mo31]{1972Mo31}  J.M. Moss, D.L. Hendrie, C. Glashausser, J. Thirion, 
\newblock Nucl.Phys. {\bf A194}, 12 (1972).
\bibitem[1972Li28]{1972Li28} A.S. Litvinenko, N.G. Shevchenko, O.Y. Buki {\it et al.}, 
\newblock Ukr. Fiz. Zh. {\bf 17}, 1197 (1972).
\bibitem[1972Ca05]{1972Ca05} J.A. Cameron, A.W. Gibb, T. Taylor, Z. Zamori, 
\newblock Can.J.Phys. {\bf 50}, 475 (1972).
\bibitem[1972Le19]{1972Le19} P.M.S. Lesser, D. Cline, P. Goode, R.N. Horoshko, 
\newblock Nucl.Phys. {\bf A190}, 597 (1972).
\bibitem[1972ArZD]{1972ArZD} R.G. Arnold, 
\newblock Thesis, Univ. Boston (1972); Diss. Abst. Int. {\bf 33B}, 1723 (1972).
\bibitem[1972Yo01]{1972Yo01} D.H. Youngblood, R.L. Kozub, J.C. Hill, 
\newblock Nucl.Phys. {\bf A183}, 197 (1972).
\bibitem[1972Ka22]{1972Ka22} D.K. Kaipov, Y.G. Kosyak, L.N. Smirin, Y.K. Shubnyi, 
\newblock Izv.Akad.Nauk SSSR, Ser.Fiz. {\bf 36}, 137 (1972); Bull.Acad.Sci.USSR, Phys.Ser. {\bf 36}, 128 (1973).
\bibitem[1971DaZM]{1971DaZM} W.G. Davies, J.S. Forster, I.M. Szoghy, D. Ward, 
\newblock AECL-3996, p.16 (1971).
\bibitem[1971Pe11]{1971Pe11} R.J. Peterson, 
\newblock Ann.Phys.(N.Y.) {\bf 65}, 125 (1971).
\bibitem[1971Sp12]{1971Sp12}  S.W. Sprague, R.G. Arns, B.J. Brunner {\it et al.}, 
\newblock Phys.Rev. {\bf C4}, 2074 (1971).
\bibitem[1971He08]{1971He08} J. Heisenberg, J.S. McCarthy, I. Sick, 
\newblock Nucl.Phys. {\bf A164}, 353 (1971).
\bibitem[1971ChZF]{1971ChZF} J. Charbonneau, N.V.De Castro Faria, J. L'Ecuyer, D. Vitoux, 
\newblock Bull. Amer. Phys. Soc. {\bf 16}, No.4, 625, JH2 (1971).
\bibitem[1970MiZQ]{1970MiZQ} W.T. Milner, F.K. McGowan, P.H. Stelson, R.L. Robinson, 
\newblock Bull.Amer.Phys.Soc. 15, No.11, 1358, DF11 (1970).
\bibitem[1970Pe15]{1970Pe15} R.J. Peterson, H. Theissen, W.J. Alston, 
\newblock Nucl.Phys. {\bf A153}, 610 (1970).
\bibitem[1970Me08]{1970Me08} F.R. Metzger, 
\newblock Nucl. Phys. {\bf A 158}, 88 (1970).
\bibitem[1970Le17]{1970Le17} P.M.S. Lesser, D. Cline, J.D. Purvis, 
\newblock Nucl. Phys. {\bf A 151}, 257 (1970).
\bibitem[1970Me18]{1970Me18} F.R. Metzger, 
\newblock Nucl. Phys. {\bf A 148}, 362 (1970).
\bibitem[1970Af04]{1970Af04} V.D. Afanasev, N.G. Afanasev, A.Y. Buki {\it et al.}, 
\newblock Yad.Fiz. {\bf 12}, 885 (1970); Sov.J.Nucl.Phys. {\bf 12}, 480 (1971).
\bibitem[1969Sp05]{1969Sp05} G.D. Sprouse, S.S. Hanna, 
\newblock Nucl.Phys. {\bf A137}, 658 (1969).
\bibitem[1969Af01]{1969Af01} V.D. Afanasev, N.G. Afanasev, I.S. Gulkarov {\it et al.}, 
\newblock Yadern. Fiz. {\bf 10}, 33 (1969); Soviet J. Nucl. Phys. {\bf 10}, 18 (1970).
\bibitem[1969Be48]{1969Be48} M.C. Bertin, N. Benczer-Koller, G.G. Seaman, J.R. MacDonald, 
\newblock Phys. Rev. {\bf 183}, 964 (1969).
\bibitem[1969Cl05]{1969Cl05} D.Cline, H.S.Gertzman, H.E.Gove {\it et al.}, 
\newblock Nucl. Phys. {\bf A 133}, 445 (1969).
\bibitem[1969To08]{1969To08} Y. Torizuka, Y. Kojima, M. Oyamada {\it et al.}, 
\newblock Phys. Rev. {\bf 185}, 1499 (1969).
\bibitem[1969Ha31]{1969Ha31} O. Hausser, T.K. Alexander, D. Pelte {\it et al.}, 
\newblock Phys. Rev. Letters {\bf 23}, 320 (1969).
\bibitem[1967Af03]{1967Af03} O.F. Afonin, A.P. Grinberg, I.K. Lemberg, I.N. Chugunov, 
\newblock Yadern.Fiz. {\bf 6}, 219 (1967); Soviet J.Nucl.Phys. {\bf 6}, 160 (1968).
\bibitem[1967Du07]{1967Du07} M.A. Duguay, C.K. Bockelman, T.H. Curtis, R.A. Eisenstein, 
\newblock Phys. Rev. {\bf 163}, 1259 (1967).
\bibitem[1967Be39]{1967Be39} R.B. Begzhanov, A.A. Islamov, 
\newblock Yadern. Fiz. {\bf 5}, 483 (1967); Soviet J. Nucl. Phys. {\bf 5}, 339 (1967).
\bibitem[1965Si02]{1965Si02} J.J. Simpson, J.A. Cookson, D. Eccleshall, M.J.L. Yates, 
\newblock Nucl.Phys. {\bf 62}, 385 (1965).
\bibitem[1965Es01]{1965Es01} M.A. Eswaran, H.E. Gove, A.E. Litherland, C. Broude, 
\newblock Nucl.Phys. {\bf 66}, 401 (1965).
\bibitem[1965Ta13]{1965Ta13} G.K. Tandon, 
\newblock Thesis, Yale University (1965).
\bibitem[1964Be32]{1964Be32} J. Bellicard, P. Barreau, D. Blum, 
\newblock Nucl. Phys. {\bf 60}, 319 (1964).
\bibitem[1964Bo22]{1964Bo22} E.C. Booth, B. Chasan, K.A. Wright, 
\newblock Nucl. Phys. {\bf 57}, 403 (1964).
\bibitem[1964El03]{1964El03} B. Elbek, H.E. Gove, B. Herskind, 
\newblock Kgl.Danske Videnskab.Selskab., Mat.-Fys.Medd. {\bf 34}, No.8 (1964).
\bibitem[1963Be29]{1963Be29} R.B. Begzhanov, A.A. Islamov, D.K. Kaipov, Y.K. Shubnyi, 
\newblock Zh.Eksperim.i Teor. Fiz. {\bf 44}, 137 (1963); Soviet Phys.JETP {\bf 17}, 94 (1963).
\bibitem[1962Be18]{1962Be18} J. Bellicard, P. Barreau, 
\newblock Nuclear Phys. {\bf 36}, 476 (1962).
\bibitem[1962St02]{1962St02} P.H. Stelson, F.K. McGowan, 
\newblock Nuclear Phys. {\bf 32}, 652 (1962).
\bibitem[1961Mc18]{1961Mc18} F.K. McGowan, P.H. Stelson, R.L. Robinson, 
\newblock Proc.Conf.Electromagnetic Lifetimes and Properties Nuclear States, Gatlinburg, Tennessee (October 1961); NAS-NRC Publ.974, p.119 (1962).
\bibitem[1961Me11]{1961Me11} F.R. Metzger, 
\newblock Nuclear Phys. {\bf 27}, 612 (1961).
\bibitem[1961Ke06]{1961Ke06} W.H. Kelly, G.B. Beard, 
\newblock Nuclear Phys. {\bf 27}, 188 (1961).
\bibitem[1961Cr01]{1961Cr01} H. Crannell, R. Helm, H. Kendall, J. Oeser, M. Yearian, 
\newblock Phys. Rev. {\bf 123}, 923 (1961).
\bibitem[1960Ad01]{1960Ad01} B.M. Adams, D. Eccleshall, M.J.L. Yates, 
\newblock Proc.Conf.Reactions between Complex Nuclei, 2nd, Gatlinbrug, A.Zucker, E.C.Halbert, F.T.Howard, Eds., John Wiley and Sons, Inc., New York, p.95 (1960).
\bibitem[1960An07]{1960An07} D.S. Andreyev, A.P. Grinberg, K.I. Erokhina, I.Kh. Lemberg, 
\newblock Nuclear Phys. {\bf 19}, 400 (1960).
\bibitem[1960An09]{1960An09} D.S. Andreyev, A.P. Grinberg, G.M.Gusinskii, K.I. Erokhina, I.Kh. Lemberg, 
\newblock Izvest.Akad.Nauk SSSR, Ser.Fiz. {\bf 24}, 1474 (1960); Columbia Tech. Transl. {\bf 24}, 1466 (1961).
\bibitem[1960Go08]{1960Go08} H.E. Gove, C. Broude, 
\newblock Proc. Conf. Reactions between Complex Nuclei, 2nd, Gatlinburg, A.Zucker, E.C.Halbert, F.T.Howard, Eds., John Wiley and Sons, Inc., New York, p.57 (1960).
\bibitem[1959Of14]{1959Of14} S.Ofer, A.Schwarzschild, 
\newblock Phys.Rev.Letters {\bf 3}, 384 (1959).
\bibitem[1959Al95]{1959Al95} D.G. Alkhazov, A.P. Grinberg, K.I. Erokhina, I.Kh. Lemberg, 
\newblock Izvest. Akad. Nauk SSSR, Ser.Fiz. {\bf 23}, 223 (1959); Columbia Tech.Transl.23, 215 (1960).
\bibitem[1959Bu12]{1959Bu12} N.A. Burgov, Y.V. Terekhov, G.E. Bizina,
\newblock Zhur. Eksptl. i Teoret. Fiz. {\bf 36}, 1612 (1959); Soviet Phys. JETP {\bf 9}, 1146 (1959).
\bibitem[1956Te26]{1956Te26} G.M. Temmer, N.P. Heydenburg, 
\newblock Phys.Rev. {\bf 104}, 967 (1956).
\bibitem[1956Me59]{1956Me59} F.R. Metzger, 
\newblock Phys.Rev. {\bf 103}, 983 (1956).
\end{theDTbibliography}

\end{document}